\documentclass{article}

\usepackage{arxiv}

\usepackage[utf8]{inputenc} 
\usepackage[T1]{fontenc}    
\usepackage{hyperref}       
\usepackage{url}            
\usepackage{booktabs}       
\usepackage{amsfonts}       
\usepackage{nicefrac}       
\usepackage{microtype}      
\usepackage{lipsum}
\usepackage{color}
\usepackage{algorithm}
\usepackage{algorithmic}
\usepackage{amsmath,amsfonts,amssymb}
\usepackage{subfigure}
\usepackage{lineno}
\usepackage{graphicx}
\usepackage{bm}
\graphicspath{{Figures/}}
\usepackage{subfigure}

\title{Image and video compression of fluid flow data}

\author{
Vishal Anatharaman, Jason Feldkamp, Kai Fukami$^*$, Kunihiko Taira\\
Department of Mechanical and Aerospace Engineering, \\
University of California, Los Angeles, CA 90095, USA\\
Corresponding author: kfukami1@g.ucla.edu
}

\begin{document}
\maketitle

\begin{abstract}
We study the compression of spatial and temporal features in fluid flow data using multimedia compression techniques. 
The efficacy of spatial compression techniques, including JPEG and JPEG2000 (JP2), and spatio-temporal video compression techniques, namely H.264, H.265, and AV1, in limiting the introduction of compression artifacts and preserving underlying flow physics are considered for laminar periodic wake around a cylinder, two-dimensional turbulence, and turbulent channel flow.
These compression techniques significantly compress flow data while maintaining dominant flow features with negligible error. 
AV1 and H.265 compressions present the best performance across a variety of canonical flow regimes and outperform traditional techniques such as proper orthogonal decomposition in some cases. 
These image and video compression algorithms are flexible, scalable, and generalizable holding potential for a wide range of applications in fluid dynamics in the context of data storage and transfer.
\end{abstract}

\section{Introduction}
\label{sec:intro}

High-fidelity simulations and experiments within the field of fluid dynamics produce exceedingly large amounts of data.  
As the need for higher fidelity simulations and advanced experimental resources expands, storage and transfer requirements for spatio-temporal data from simulations become a major challenge.
To address this issue, spatio-temporal redundancies or repeated dominant flow features can be exploited by a variety of compression techniques to alleviate memory constraints for fluid flow data storage.
A variety of compression techniques, including modal analysis~\cite{Holmes,Schmid2010,taira17}, sub-sampling and local re-simulation~\cite{wu20}, and deep learning~\cite{liu18,glaws2020deep,mohan2020spatio,momenifar2022dimension} have been considered in an effort to reduce the size of fluid flow data. 
Although effective, these techniques can be application-specific and struggle to achieve substantial compression ratios without introducing undesirable compression artifacts such as discontinuities or deletions of flow features.

In comparison, multimedia compression techniques are general and simple to use, and have benefited from demand for the modern technologies of high-resolution video streaming~\cite{apostolopoulos2002video,rao2011network,jiang2021survey} and video-conferencing~\cite{egido1988video,augestad2009overcoming,mpungose2021lecturers}. 
These compression techniques are classified into two groups: lossless compression and lossy compression~\cite{said1996image}.
With lossless techniques, the data retrieved from or reconstructed from the compressed state is identical to that preceding the application of a compression algorithm.
Hence, this is preferred for archival purposes and used for medical imaging~\cite{liu2017current} and technical drawings~\cite{arora2014comprehensive}.
In contrast, processed data with lossy techniques do not necessarily match the original data, enabling a significant data-size reduction in the compressed state.
Since this may introduce compression artifacts such as discontinuities in image data or the loss of high spatial frequency information, it is suitable for natural images such as photographs in applications where imperceptible loss may be acceptable~\cite{guo2016building}. 
We consider here the impacts of such losses on fluid mechanics simulation data to assess the costs of applying lossy techniques.
In 2003, Schmalzl~\cite{schmalzl2003using} considered multimedia data compression for fluid flows with an example of Rayleigh-B\'{e}nard convection.
With multimedia compression technologies having undergone significant advances in the last two decades, we reassess image and video compressions with modern algorithms for applications to fluid flow data.

Lossy techniques of interest typically involve frequency-domain transformation, filtering, and entropy coding as components in the compression process. The development of the discrete cosine transform (DCT)~\cite{ahmed1974discrete,ahmed1991came} has played a crucial role in image compression, and is the basis of Joint Photographic Experts Group (JPEG)~\cite{mitchell1992digital}.
The emergence of JPEG enabled efficient image compression in a wide range of communities and it became a generally accepted format for digital images.
After the development of DCT, wavelet transforms began to be utilized for image compression in such algorithms as JPEG2000 (JP2)~\cite{taubman2012jpeg2000}, which achieves better compression than the DCT of JPEG as a result of multi-scale properties of wavelets.
{It is worth pointing out that there have been studies on computational fluid dynamics that leverage wavelet methods to efficiently decompose multi-scale features for applications in turbulence modeling and simulations~\cite{schneider2010wavelet}.}

In tandem with the growth of image compression techniques, advancement in video compression technologies followed suit since video data can be characterized as a time series of image frames.
Generally, these time frames include both spatial and temporal redundancies.
In fact, we often see the similarities (redundancies) between temporally adjacent frames or spatially adjacent pixels.
Video compression algorithms are designed to remove such redundancies and obtain a compact form of the original information.
Current video compression technologies are generally based on the DCT~\cite{hoffman2012data}.
Although other candidates including fractal compression~\cite{fisher1994fractal,fisher2012fractal}, matching pursuit~\cite{mallat1993matching}, and discrete wavelet transform (DWT) have been investigated as the subject of some studies, these are still not used in practical products.
Moving Picture Experts Group (MPEG) series have been traditionally used for video compression of high-definition television~\cite{tudor1995mpeg,haskell1996digital,bosi1997iso}.
H.2xx series was then developed and they have achieved significant compression compared to the conventional MPEGs~\cite{wiegand03,pastuszak2015algorithm}.
Especially in the recent versions such as H.264 and H.265, motion compensation, quantization, and entropy coding are applied for efficient video compression.
More recently, AOMedia Video 1 (AV1), an open, royalty-free video coding format, was released in 2018, achieving enhanced compression compared to the aforementioned techniques~\cite{chen18,han2021technical}.

To meet the demand for these image and video compression tools, significant investment and research have produced compression techniques of impressive efficiency and usability in addition to free video encoders~\cite{ffmpeg21} to promote widespread accessibility.
As such, leveraging these multimedia-inspired compression techniques should also be of particular interest to the fluid dynamics community given the massive scale of data produced, stored, and transferred.
A standardization on one or more multimedia compression formats for storing fluid flow data in a compressed representation can yield dividends in research output by allowing greater access to high-fidelity fluid flow data sets and by removing memory constraints as a barrier to entry.

\begin{figure}
  \centerline{
  \includegraphics[width=0.9\textwidth]{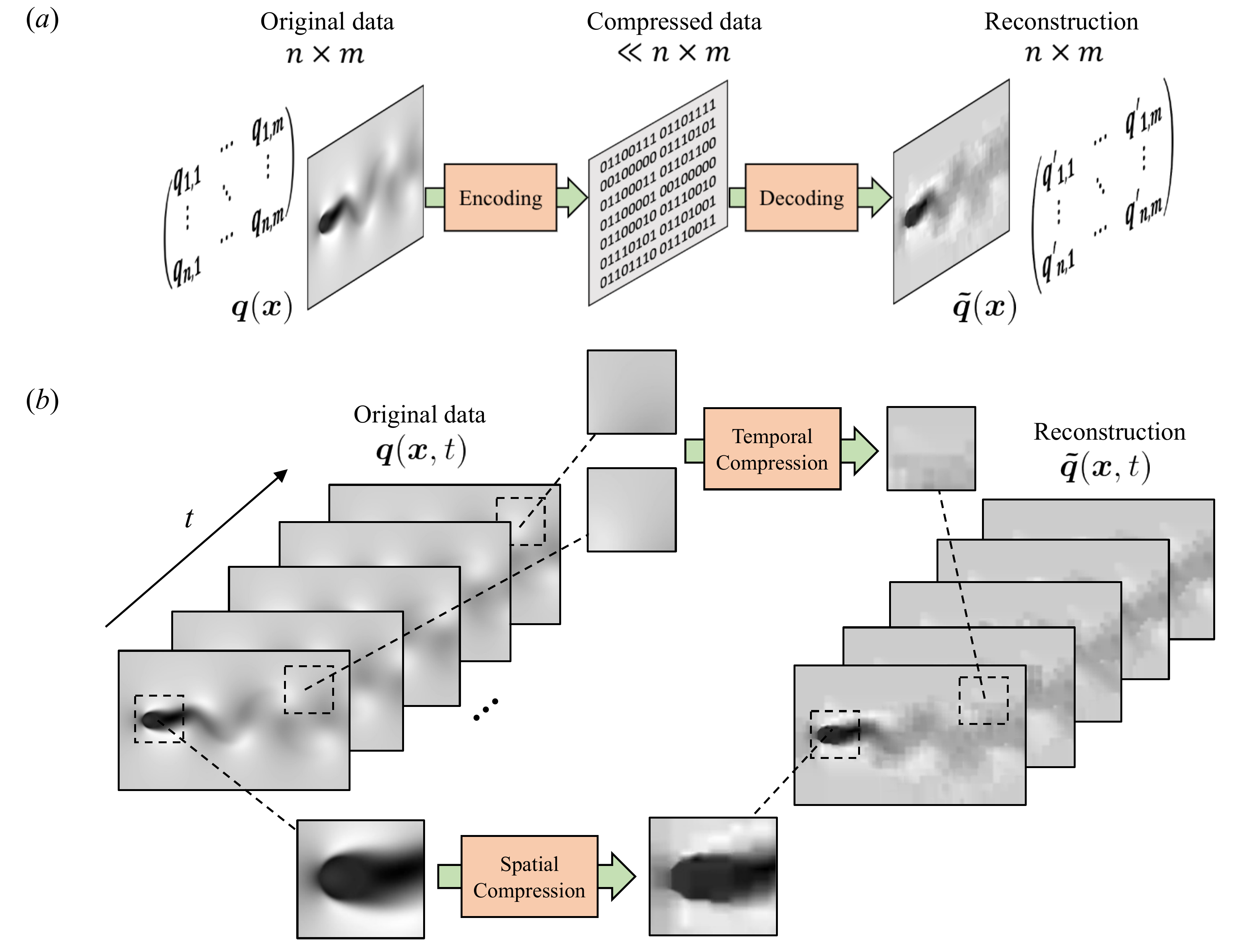}
  }
  \caption{
  $(a)$ Spatial compression: an example velocity field of flow over a cylinder ${\bm q}({\bm x})$ is represented as a grayscale image, encoded using an image-based technique to a compressed form, and reconstructed as ${\bm{\tilde{q}}}({\bm x})$ using a decoder. 
  $(b)$ Spatio-temporal compression: multiple snapshots of this flow field data ${\bm q}({\bm x},t)$ are represented as a grayscale video and are compressed to ${\bm{\tilde{q}}}({\bm x},t)$ with both spatial and temporal techniques.  
  }
\label{fig2}
\end{figure}

This paper investigates the effectiveness of these image and video compression techniques on fluid flow data.
Spatial image compression techniques, such as JPEG and JP2, alongside spatio-temporal video compression techniques, namely H.264, H.265, and AV1, are examined for various flow fields, including laminar cylinder flow, two-dimensional turbulence, and turbulent channel flow.
Field variables from simulation data, such as streamwise velocity and vorticity, are represented as grayscale images, and multiple snapshots are packaged into a video. 
These videos are then encoded into a compressed form using the aforementioned multimedia compression methods.
Modern techniques can compress flow data well below 10\% of the original file size with negligible error and preserve the underlying physics of the flow. 
Although this paper focuses on applications to canonical fluid flows, the flexibility and scalability of these algorithms suggest an expansive potential within this field.

Compression is a process in which data is compressed (encoded) into a representation that uses less data, and decompressed (decoded) into identical data in the case of lossless compression or nearly-identical data in the case of lossy compression.
Through this procedure, a compression method reduces bits of the original data ${\bm q}({\bm x},t)$ by eliminating statistical redundancies that may be contained within temporally adjacent frames and spatially adjacent pixels.
{In general, a data compression algorithm is referred to as an encoder $\phi$ while one that performs the decompression is called a decoder~$\psi$,
\begin{align}
    {\bm \gamma}({\bm x},t) = \phi({\bm q}({\bm x},t)),~~~{\bm q}({\bm x},t)\approx \tilde{\bm q}({\bm x},t)=\psi({\bm \gamma}({\bm x},t)),
\end{align}
where ${\bm \gamma}({\bm x},t)\in\mathbb{R}^{m}$ is the compressed data corresponding to the original data ${\bm q}({\bm x},t)\in\mathbb{R}^{n}$ with $m\ll n$.
Depending on the extent of compression, the data, and a choice of encoder/decoder, the reconstruction $\tilde{\bm q}({\bm x},t)\in\mathbb{R}^{n}$ generally includes some amount of error.}

The data compression process is illustrated in figure~\ref{fig2} for both image and video compressions.
Figure~\ref{fig2}$(a)$ depicts a lossy spatial image compression technique, involving quantization of the image data in a compressed space and producing a reconstruction in the image space showing the operations of JPEG and JP2. 
Figure~\ref{fig2}$(b)$ provides a visualization of a spatio-temporal compression technique, exploiting a redundant block of a frame that remains consistent across subsequent frames, similar to H.264, H.265, and AV1.
As these algorithms originated in the multimedia industry, they are optimized for human viewers and involve the removal of high-frequency components in the data and down-sampling of the color spectrum such that the eyes cannot easily distinguish compressed data from the original data.  
For the purposes of this study, we only consider grayscale images and videos, which are comprised only of a single-component field data matrix, denoted as ${\bm{\tilde{q}}}({\bm x})$.  
This is in contrast to full-color data, which requires red, green, and blue components, and is unnecessary for the current analysis as we are interested in considering field variables individually.
Herein, we consider the application of five compression techniques on grayscale images and videos. 
The encoding schemes, which package the data into a compressed binary form, are detailed in what follows.

\section{Compression techniques}
\subsection{Image Compression}
\subsubsection{JPEG}

\begin{figure}
  \includegraphics[width=1\textwidth]{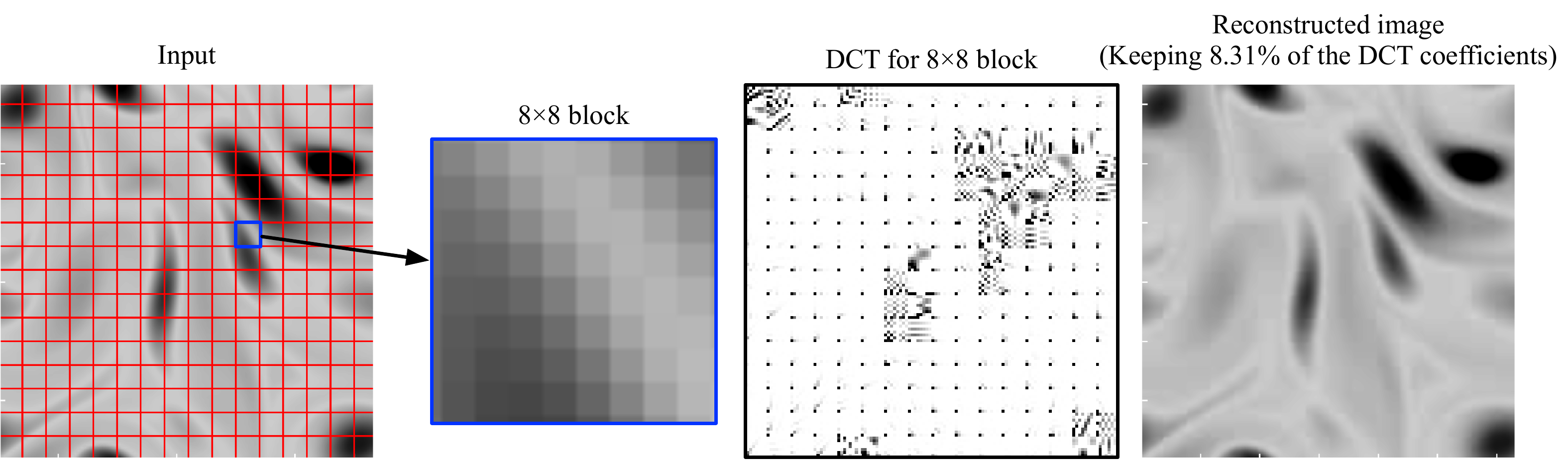}
  \caption{JPEG compression process with an example of two-dimensional isotropic turbulent vorticity.}
\label{fig:dct}
\end{figure}

Let us first describe JPEG, which is a standard lossy spatial compression used for encoding image data based on the discrete cosine transform (DCT){~\cite{ahmed1974discrete}}. 
An example of a JPEG compression process with a vorticity field of two-dimensional decaying isotropic turbulence is presented in figure~\ref{fig:dct}.
The images are partitioned into $8 \times 8$ blocks in a left-to-right, top-to-bottom scan.
Pixel values within blocks are quantized to values of $[-128, 127]$ from $[0, 255]$.
The forward DCT is individually performed at each block and outputs compressed data.
The DCT for $8 \times 8$ blocks is mathematically expressed as
\begin{align}
    &F(k_x,k_y) = \frac{1}{4}C(k_x)C(k_y)\biggl(\sum_{i_x=0}^7\sum_{i_y=0}^7f(i_x,i_y) \cos\biggl(\frac{(2i_x+1)k_x\pi}{16}\biggr) \cos\biggl(\frac{(2i_y+1)k_y\pi}{16}\biggr)\biggr),\\
    &f(i_x,i_y) = \frac{1}{4}\biggl(\sum_{k_x=0}^7\sum_{k_y=0}^7C(k_x)C(k_y)F(k_x,k_y) \cos\biggl(\frac{(2i_x+1)k_x\pi}{16}\biggr) \cos\biggl(\frac{(2i_y+1)k_y\pi}{16}\biggr)\biggr),
\end{align}
where
\begin{equation}
C(k) = 
    \begin{cases}
        1/{\sqrt{2}} & {\rm for}~k=0 \\
        1            & {\rm otherwise.}
    \end{cases}
\end{equation}
Here, $F(k_x,k_y)$ denotes the DCT coefficient corresponding to the horizontal wavelength $k_x$ and vertical wavelength $k_y$ and $f(i_x,i_y)$ describes the pixel value at the location corresponding to $i_x$ and $i_y$.
In other words, the forward DCT takes as input a discrete signal of 64 points and produces coefficients for a linear combination of 64 unique basis signals, each denoting a specific spatial wavelength.
Most of the spatial domain information is concentrated across lower wavelength because of slow spatial variation from one pixel to the next in image data. 
This quality permits lossy quantization, which refers to constant values in a quantization table $Q(k_x,k_y)$ with 64 elements.
The DCT coefficient is normalized by a constant $Q(k_x,k_y)$ in an element-wise manner,
\begin{align}
     F^Q(k_x,k_y) = \lfloor \dfrac{F(k_x,k_y)}{Q(k_x,k_y)}\rfloor
\end{align}
where $F^Q(k_x,k_y)$ is a normalized coefficient and the operation $\lfloor\cdot \rfloor$ denotes rounding to the nearest integer.
Quantization tables are provided by the Joint Photographics Experts Group.
Note that dividing the DCT coefficients by values in the quantization table reduces high-wavenumber coefficients to $0$, which permits efficient entropy coding (explained later) to perform the cutoff at high frequencies~\cite{wallace92}. 
The resulting quantized DCT coefficients form a matrix of size 8 $\times$ 8 with low-wavenumber components generally located in the top-left of the matrix and high-wavenumber coefficients at the bottom-right, as a consequence of the similar distribution of spatial modes to which these coefficients correspond.
Subsequently, quantized coefficients are ordered from low to high wavenumbers.

To reduce the data size, entropy coding~\cite{sze2014entropy,duda2015use}, a lossless method of compressing bitstreams with redundancies, is then performed for the output of DCT.
The idea of entropy coding is used not only for JPEG but also other image/video compression techniques such as JP2, H.2xx series, and AV1.
To express the encoding-based data compression, let us consider a message of {\sf DAEBCBACBBBC} (12 characters).
Since this message includes five different characters, it needs to prepare 3 bits to convert these characters to bits or binary digits representation.
Here, we use the following conversion table,
\begin{table}[H]
\begin{center}
\begin{tabular}{ccccc}
{\sf A}   & {\sf B} & {\sf C} & {\sf D} & {\sf E}   \\
000 & 001 & 010 & 011 & 100
\end{tabular}
\end{center}
\end{table}
With this table, the message is expressed as
\begin{table}[H]
\begin{center}
\begin{tabular}{cccccccccccc}
{\sf D} & {\sf A} & {\sf E} & {\sf B} & {\sf C} & {\sf B} & {\sf A} & {\sf C} & {\sf B} & {\sf B} & {\sf B} & {\sf C} \\
011 & 000 & 100 & 001 & 010 & 001 & 000 & 010 & 001 & 001 & 001 & 010
\end{tabular}
\end{center}
\end{table}
As shown, the number of bits is 36.
The idea of the encoding-based compression is to prepare an adaptive conversion table assigning a shorter bit length for characters that appear in a high probability and a longer bit length for characters that barely appear. 
For example, the following adaptive table can be used: 
\begin{table}[H]
\begin{center}
\begin{tabular}{ccccc}
{\sf A}   & {\sf B} & {\sf C} & {\sf D} & {\sf E}   \\
110 & 0 & 10 & 1110 & 1111
\end{tabular}
\end{center}
\end{table}
With this new table, the message can be expressed as
\begin{table}[H]
\begin{center}
\begin{tabular}{cccccccccccc}
{\sf D} & {\sf A} & {\sf E} & {\sf B} & {\sf C} & {\sf B} & {\sf A} & {\sf C} & {\sf B} & {\sf B} & {\sf B} & {\sf C} \\
1110 & 110 & 1111 & 0 & 10 & 0 & 110 & 10 & 0 & 0 & 0 & 10
\end{tabular}
\end{center}
\end{table}
The current table can save the total number of bits for the message from 36 to 25.
Modern data compression techniques efficiently find such an adaptive conversion table for saving image and video sizes.

The presence of a better adaptive conversion can be proven with the source coding theorem~\cite{shannon1948mathematical}.
For any data, the expected code length should satisfy the relationship,
\begin{align}
{\displaystyle \mathbb {E} _{\beta\sim P}[l(d(\beta))]\geq \mathbb {E} _{\beta\sim P}[-\log _{b}(P(\beta))]}, 
\end{align}
where $l$ is the number of symbols in a message, $d$ is the coding function, $b$ is the number of symbols in a table, and $P$ is the probability of the original symbol. 
An entropy coding method attempts to approach the lower bound.
For JPEG compression, Huffman coding~\cite{huffman1952method} is used to determine an adaptable table composed of the estimated probability of occurrence for each possible value.
Huffman coding uses binary trees~\cite{knuth1973fundamental} for efficient encoding.

\subsubsection{JPEG2000 (JP2)}

\begin{figure}[t]
  \centerline{
  \includegraphics[width=1\textwidth]{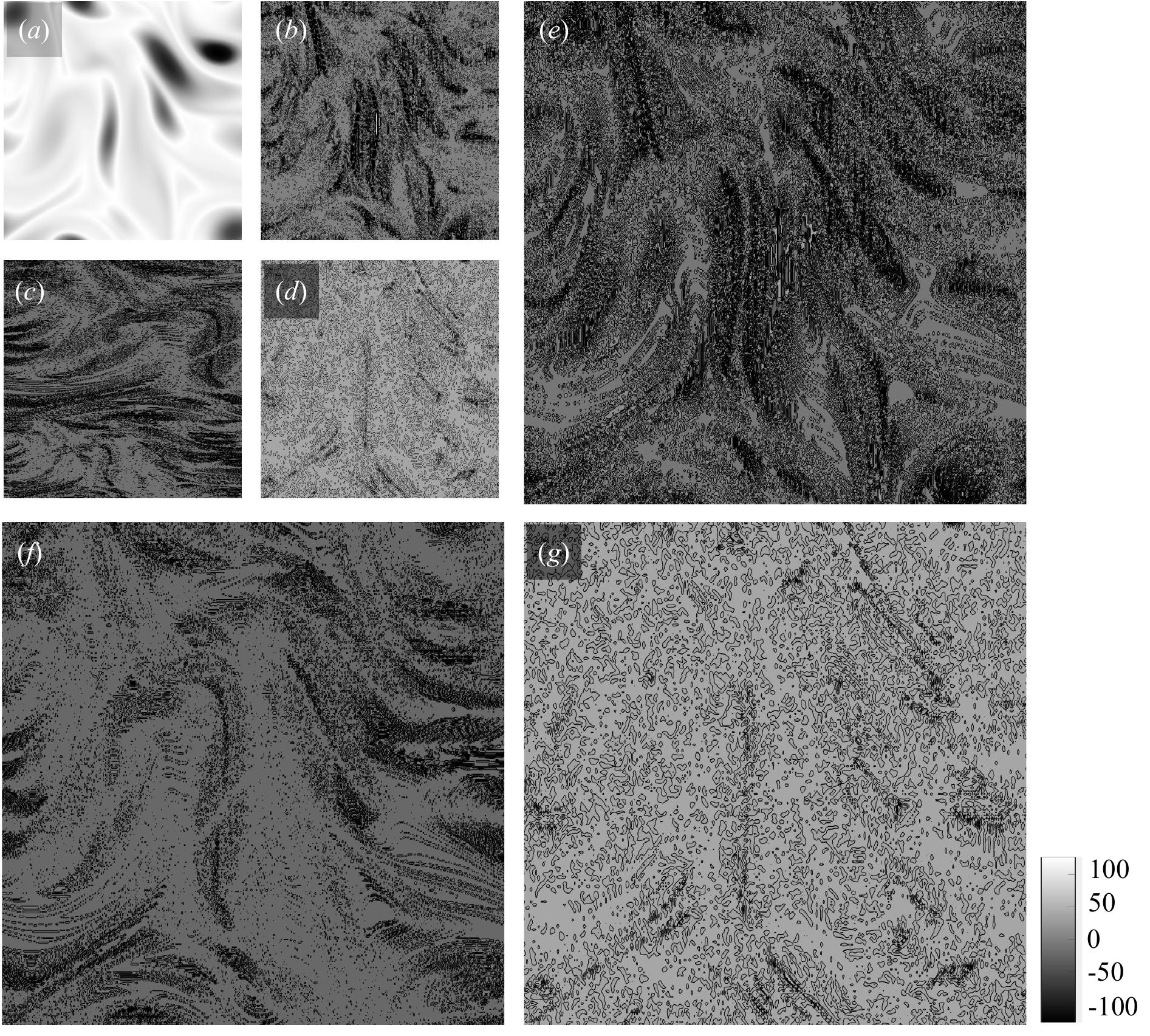}
  }
  \caption{
  An example of the two-level discrete wavelet transform for two-dimensional homogeneous turbulent vorticity field, used in JP2 compression. 
  High-pass filtering yields three large images.
  Low-pass filtering and downscaling are then performed, producing the three small images.
  $(a)$ The final approximation image.
  $(b,e)$ Vertical, $(c,f)$ horizontal, and $(d,g)$ diagonal coefficients of the second $(b-d)$ and the first levels $(e-g)$ are shown.
  } 
\label{fig:dwt}
\end{figure}

JPEG2000 (JP2) is a successor to JPEG. JP2 operates using a similar four-step process to JPEG, involving image partitioning, frequency-domain transformation, quantization, and entropy coding.
In contrast to JPEG, JP2 introduces more advanced dynamic tiling algorithms based on variable-sized macroblocks.
This makes use of {discrete wavelet transform (DWT)~\cite{daubechies1988orthonormal,mallat1989multiresolution}}, and performs additional preprocessing prior to entropy coding. 
Tiling in this context refers to the partitioning of the source image into several non-overlapping rectangular blocks, each of which is processed distinctly. 
Whereas JPEG restricts tile sizes to 8 $\times$ 8, tiles in JP2 can be of arbitrary size up to the image dimensions.
The DWT is applied to each tile in a manner similar to DCT, decomposing a signal into a linear combination of wavelet functions.
The coefficients in this linear combination correspond to a specific wavelet basis function in the signal.
A wavelet can be defined as a scale and shift of {a basis wavelet~\cite{ricker1953wavelet}}.
Child wavelets~\cite{heil1989continuous} are generally considered for DWT, given by
\begin{align}
    \psi_{g,r}(s) = \dfrac{1}{2^{g/2}} \psi \biggl(\dfrac{s -  2^gr}{2^g}\biggr)
\end{align}
where $g$ is a scaling factor, $r$ is a shift factor, and $s$ corresponds to the index of the one-dimensional representation of an image. 
In other words, the flow field snapshot is converted to a one-dimensional representation and the independent variable upon which this one-dimensional signal $f(s)$.
The DWT coefficient given a wavelet of the preceding definition is then
\begin{align}
    F_{g,r} = \int_{-\infty}^{\infty} f(s) \psi_{g,r} ds
\end{align}
where $f(s)$ is a one-dimensional signal.
The signal can be reconstructed through the summation of the product of each coefficient with the corresponding wavelet.
In a discrete interpretation, this is written as
\begin{align}
    f(s) = \sum_{g = -\infty}^{\infty} \sum_{r = -\infty}^{\infty} F_{g,r} \psi_{g,r}.
\end{align}
The summation bounds for both the calculation of the coefficients and reconstruction of the signal can be set to finite values and can still produce lossless reconstructions assuming that the wavelets contain the maximum and minimum wavelengths within the source image.

An example of the DWT for the vorticity field of two-dimensional decaying turbulence is presented in figure~\ref{fig:dwt}.
The DWT can be applied recursively to one-dimensional signals to produce higher fidelity representations of data. 
As such, successive high-pass filters are applied on down-sampled images, producing a higher fidelity representation of spatial frequencies in the source image. 
The DWT can be extended to higher dimensions by applying the one-dimensional DWT on rows and columns. 
The recursive application of the DWT produces $2^n$ distinct filtered images where $n$ is the number of times the DWT is applied.

Similar to JPEG compression, the DWT coefficients are quantized following the transformation on the wavespace,
\begin{align}
    F^Q_{a,b} = {\rm sign}(F_{a,b}) \lfloor \frac{|F_{a,b}|}{\Delta_b} \rfloor,
\end{align}
where $\Delta_b$ is defined as the quantization step. 
DWT coefficients within the range $(-\Delta_b, \Delta_b)$ are quantized to $0$. 
Following quantization, the coefficients are processed in preparation for entropy coding.
Arithmetic coding~\cite{rissanen1979arithmetic} is used for entropy coding in JPEG2000.
While Huffman coding separates the original data into component symbols and replaces each with a code in a table, Arithmetic coding encodes the entire message into a single number represented with an arbitrary-precision fraction $p_a$, where $0 \leq p_a < 1$.

\subsection{Video Compression}

\subsubsection{H.264 (AVC)}

The H.264 video compression includes a multi-step process, consisting generally of prediction, transformation (a set of frequency-domain representation and quantization in image compression), and entropy encoding on the encoder side.
A similar process for file reconstruction is performed on the decoder side.
Prediction in video compression amounts to an operation to remove redundancies in the given signal.
H.264 supports a range of prediction options such as intra-prediction used for data within the current frame, inter-prediction for motion compensation, and multiple block-size-based predictions.
An accurate prediction implies that the residual contains very little information, amounting to good compression performance.

Video data is first partitioned into macroblocks of dimension 16 $\times$ 16 pixels.
A prediction of the current macroblock is formed using 4 $\times$ 4 and 16 $\times$ 16-sized blocks in the case of intra-frame prediction, referring to predictions from surrounding blocks within the same frame.
A range of block sizes from 4 $\times$ 4 to 16 $\times$ 16 are also considered in the case of inter-frame prediction, referring to predictions from previously coded frames.
Macroblock prediction is further discretized into intra-prediction with neighboring blocks in the current frame, blocks in a previously coded frame, and blocks from up to two previously coded frames.

\begin{figure}
  \centerline{
  \includegraphics[width=1\textwidth]{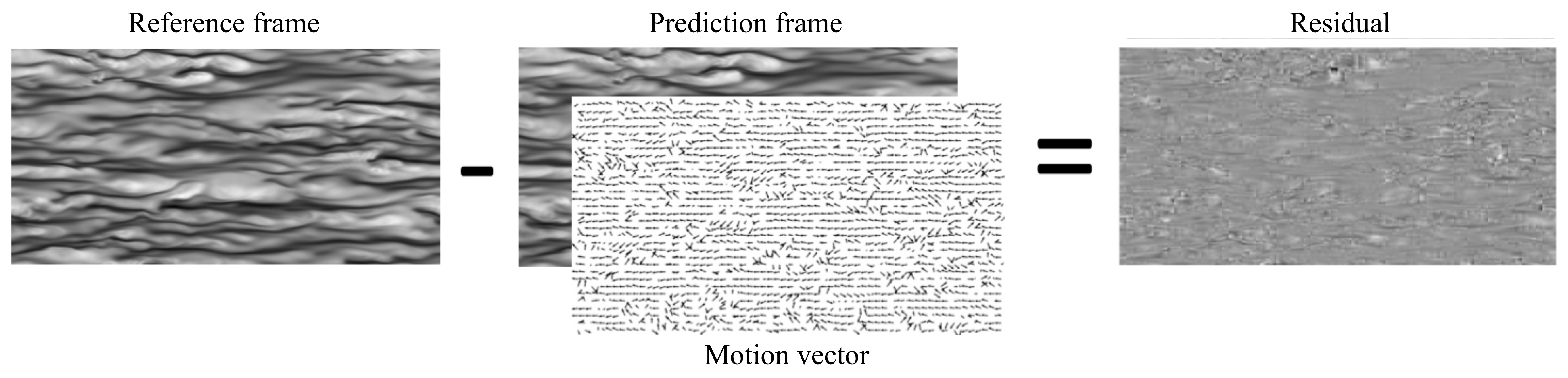}
  }
  \caption{Motion compensation (MC), used in temporal compression of H.264, H.265, and AV1, is exemplified by the calculation of a motion vector field, describing the translation of pixels between successive frames. 
  As an example, a streamwise velocity field $u$ of three-dimensional turbulent channel flow is considered.
  Subtracting {the prediction frame} from {the reference frame} yields a residual image, which is stored.
   }
\label{fig:motioncompensation}
\end{figure}

In intra-prediction, the size of prediction macroblocks can be three cases: 16 $\times$ 16, 8 $\times$ 8, or 4 $\times$ 4 pixels~\cite{wiegand03}. 
The choice of block size is made primarily based on prediction efficiency.
One of several prediction modes where each prediction mode indicates a direction in which to extrapolate pixel values or to average across all pixels. 
In the case of inter-prediction, the reference frame, where the prediction block is situated, is chosen from several previously decoded frames.
{A motion vector is obtained for the current macroblock based on the offset from the prediction block or from previously coded motion vectors.
The prediction frame is then calculated using the motion vector and the reference frame.
The residual between the reference and prediction frames is finally stored, as illustrated in figure~\ref{fig:motioncompensation}.}
These motion vectors are optionally weighted to account for temporal proximity between frames, and are sent into the data stream. 
Generally, a deblocking filter is further applied to each frame to store in a decoded format for subsequent inter-frame predictions in smoothing the sharp edges caused by the use of block coding~\cite{list2003adaptive}.

Transformation involves the conversion of blocks to frequency-domain representations and quantization of coefficients corresponding to high wavelength data.
The DCT step is given by the transformation of block $X$ by matrix $A$ into DCT coefficients $Y$ for each macroblock.
For the case of 4 $\times$ 4 blocks, these matrices $X,Y\in \mathbb{R}^{4\times 4}$ are expressed as
\begin{equation}
    Y = AXA^T = \begin{bmatrix}
        a & a & a & a\\
        b & c & -c & -b\\
        a & -a & -a & a\\
        c & -b & b & -c
    \end{bmatrix}X\begin{bmatrix}
        a & b & a & c\\
        a & c & -a & -b\\
        a & -c & -a & b\\
        a & -b & a & -c
    \end{bmatrix},
    \label{eq:axa}
\end{equation}
where
     $a={1}/{2}$,
     $b=\sqrt{{1}/{2}}\cos({\pi}/{8})$,
     $c=\sqrt{{1}/{2}}\cos({3\pi}/{8})$.
The rows of $A$ are orthonormal.
To calculate equation~\ref{eq:axa} on a practical processor, the approximation for $b$ and $c$ is required. 
This is achieved with a fixed-point approximation, which is equivalent to scaling each row of $A$ by 2.5 and rounding to the nearest integer.
A core transform $C_{f4}$, which scales each term of $A$ by 2.5 and rounds to the nearest integer, and a scaling matrix $S_{f4}$, which restores norms of row of $C_{f4}$ to 1 by scaling, are respectively defined as,
\begin{equation}
    C_{f4} = \begin{bmatrix}
        c_1 & c_1 & c_1 & c_1\\
        c_2 & c_1 & -c_1 & -c_2\\
        c_1 & -c_1 & -c_1 & c_1\\
        c_1 & -c_2 & c_2 & -c_1
    \end{bmatrix},
    \quad 
    S_{f4} = \begin{bmatrix}
        s_1 & s_2 & s_1 & s_2\\
        s_2 & s_3 & s_2 & s_3\\
        s_1 & s_2 & s_1 & s_2\\
        s_2 & s_3 & s_2 & s_3\\
    \end{bmatrix},
\end{equation}
where $c_1 = 1$, $c_2=2$, $s_1=1/4$, $s_2=1/(2\sqrt{10})$, and $s_3=1/10$.
Matrix $Y$ is then determined as
\begin{align}
    Y = [C_{f4}XC_{f4}^T] S_{f4}.
\end{align} 
Similar DCT approximations are specified for other block sizes, involving $C_{f8}$, $S_{f8}$, and others~\cite{wiegand03}.
A quantization mechanism similar to JPEG is applied, with a quantization table specified for various block sizes. 
The quantized DCT coefficients are then traversed in an oscillating, ``zig-zag'' manner from low- to high-wavenumber components.
Entropy coding is finally applied to the output of DCT.


\subsubsection{H.265 (HEVC)}

H.265 was released in 2013 as the successor to H.264. 
While the fundamental architecture is unchanged from H.264, H.265 makes use of coding tree units which are similar to macroblocks but expand the range of possible dimensions, with variable dimensions selected by the encoder, allowing coding tree units to be divided into sub-blocks. 
Predictions and reconstructions are performed on coding tree units and supported sub-block sizes range from 64 $\times$ 64 to 4 $\times$ 4 pixels. 
Motion vectors are predicted based on those of adjacent units or blocks in the case of intra prediction, or previous encoded frames in the case of inter prediction. 
As in H.264, the DCT is performed at the coding tree block level, and the resultant coefficients are subjected to scalar quantization and entropy coding. 
Instead of deblocking filter in H.264, a sample adaptive offset filter is applied within the prediction loop to improve the quality of the compressed data~\cite{sullivan12}.

\subsubsection{AV1}

AV1 was released in 2018 in an effort to replace the H.2XX series of video compression algorithms.
In AV1, much of the fundamental architecture from H.2XX is maintained.
Frames are partitioned using a 4-way partition tree with dimensions ranging from 128 $\times$ 128 to 4 $\times$ 4. 
AV1 supports 56 directional spatial modes for intra-frame prediction with finer angle variations than that provided by H.2XX.
AV1 extends the number of reference frames that any given frame can use to perform predictions to seven references for inter-frame prediction, thus enabling more accurate encoding of data with rich temporal characteristics.
Motion vector field formation is improved by expanding the spatial search domain for vector candidates and through the utilization of a temporal motion field estimation system.
AV1 also extends frequency-domain transform algorithms to include the asymmetric discrete sine transform with a richer set of transform kernels for varying block sizes. 
Entropy coding and scalar quantization are also used as well as H.2XX compression algorithms.
To perform deblocking, a constrained directional enhancement filter and loop restoration filters are applied~\cite{chen18}.
These filters are able to effectively remove artifacts without causing blurring, compared to conventional deblocking filters~\cite{midtskogen2018av1}.

\section {Flow Fields} 

Let us apply the compression techniques presented above to representative fluid flow data sets from numerical simulations.
We describe herein the problem setup of the example flow fields we analyze and the simulation approach used to generate them.

\subsection{Two-dimensional laminar cylinder wake}

Bluff body flow forms a large class of problems, such as the vortex shedding around a cylinder.
We first apply the compression techniques to the two-dimensional cylinder wake obtained by direct numerical simulation (DNS)~\cite{TC2007,CT2008}.
The governing equations are the incompressible Navier--Stokes equations,
\begin{eqnarray}
\nabla \cdot {\bm u}= 0, \\
\frac{\partial {\bm u}}{\partial t}+{\bm u} \cdot \nabla {\bm u}=-\nabla p+\frac{1}{Re_D}\nabla^2\bm u,
\end{eqnarray}
where $\bm u$ and $p$ are the non-dimensionalized velocity vector and pressure, respectively.  
All variables are non-dimensionalized with the fluid density $\rho$, the uniform velocity $U_\infty$, and the cylinder diameter $D$.
The Reynolds number is defined as $Re_D=U_\infty D/\nu=100$ with $\nu$ being the kinematic viscosity.
We consider five nested levels of multi-domains with the finest grid level covering $(x, y) = [-1, 15] \times [-8, 18]$ and the largest domain being $(x,y) = [-5, 75] \times [-40, 40]$. 
The time step for DNS is $\Delta t=2.50\times 10^{-3}$. 
We extract the domain around a cylinder body over $(x^*,y^*)/D = [-0.7, 15] \times [-5, 5]$ with $(N_x,N_y) = (500, 300)$ and $(\Delta x,\Delta y) = (0.0314, 0.0333)$.
The flow exhibits vortex shedding with a single period with 21 snapshots.
For the compression analysis, 160 temporal snapshots of grayscale images of the streamwise velocity field $u$ are considered.

\subsection{Two-dimensional decaying homogeneous isotropic turbulence}

To examine the data compression performance by the present techniques for more complex turbulent flows, we also consider a two-dimensional decaying homogeneous isotropic turbulence.
This time-varying flow can be regarded as a canonical fluid flow example for a broad range of turbulent flows.
The data set is prepared by DNS using the two-dimensional vorticity transport equation~\cite{TNB2016}.
We set the initial Reynolds number $Re_0\equiv u^*l_0^*/\nu=80.4$, where $u^*$ is the characteristic velocity obtained by the square root of the spatially averaged initial kinetic energy, {$l_0^*=[2{\overline{u^2}}(t_0)/{\overline{\omega^2}}(t_0)]^{1/2}$} is the initial integral length, and $\nu$ is the kinematic viscosity.
The computational domain and the numbers of grid points  are set to $L_x=L_y=1$ and $N_x=N_y=128$, respectively.
We use 1000 snapshots in an eddy turn-overtime of $t\in[2,6]$ with a time interval of $\Delta t=0.004$.
For the data compression analysis, $128 \times 128$ grid with grayscale contours of the vorticity field $\omega$ are used.

\subsection{Turbulent channel flow}

To further demonstrate the present data compression techniques, we also examine turbulent channel flow at a friction Reynolds number of $\displaystyle{{Re}_\tau = u_\tau  \delta/\nu}=180$, where $u_\tau$ is the friction velocity, $\delta$ is the half-width of the channel, and $\nu$ is the kinematic viscosity.
This flow involves a broader range of spatio-temporal flow scales and fewer redundancies compared to the previous two examples.
The data sets are prepared by a three-dimensional DNS~\cite{FKK2006,FFT2020b}, numerically solving the incompressible Navier--Stokes equations.
The present DNS has been validated by comparison with spectral DNS data of Moser et al~\cite{moser1999direct}.
The streamwise, wall-normal, and spanwise spatial coordinates are denoted by $x$, $y$, and $z$, respectively.
The size of the computational domain and the number of grid points here are $(L_{x}, L_{y}, L_{z}) = (4\pi\delta, 2\delta, 2\pi\delta)$ and $(N_{x}, N_{y}, N_{z}) = (256, 96, 256)$, respectively.  
The grids in the streamwise and spanwise directions are taken to be uniform, while that in the $y$ direction is a non-uniform grid.
The no-slip boundary condition is imposed on the walls and a periodic boundary condition is applied to the $x$ and $z$ directions.
The flow is driven by a constant pressure gradient.  
In what follows, we denote wall-unit quantities with the superscript $+$.

For the present study, an $x-z$ cross-sectional streamwise velocity $u$ at $y^+=13.2$ is analyzed, where representative streak structures are present~\cite{KMM1987}.
Fifty temporal snapshots of a $256 \times 256$ spatial grid over {$t^+\in[0,63]$} are formatted into grayscale data and are used for compression assessment.

\section {Results}

For image compression, matrices of flow field data corresponding to specific temporal snapshots are represented as uncompressed grayscale images 
{with 8 bits. 
We have confirmed that the loss of precision in converting to the grayscale image does not significantly affect the statistics of fluid flows.}
These images are compressed using JPEG and JP2 encoders.
For video compression, the matrices of flow field data are represented as grayscale images and concatenated into uncompressed, grayscale videos.
This raw video file is used as the input to the H.264, H.265, and AV1 encoders.
To obtain control over the outputted file size for the purposes of this analysis, a two-pass encoding scheme is considered.
In the two-pass encoding, the encoder runs twice.
The first run is used to collect some information and statistics such as how many bytes would be needed for data compression and the second run performs the actual encoding.
These two processes enable the use of the information collected in the first run to achieve enhanced compression.

This study uses FFmpeg~\cite{ffmpeg21}, a free and open-source software consisting of various libraries for handling video, audio, and other multimedia files. 
These libraries can be easily used from the command line {\texttt{ffmpeg}}.
Default encoding settings for the relevant FFMpeg library are used to maintain consistency across all tests.

\subsection {Image compression}

To establish a baseline performance for comparison, individual flow snapshots are compressed using {singular value decomposition (SVD)~\cite{stewart1993early}}.
Individual snapshots are decomposed into left and right singular vector matrices $U$ and $V^T$, and a diagonal matrix $\Sigma$ containing the singular values. 
Snapshots are reconstructed by retaining the $r$ leading modes.
Here, the compression ratio $\eta$ in the SVD context is then defined as
\begin{align}
    \eta = \frac{r(m+n+1)}{mn},
\end{align}
where $m$ and $n$ are the snapshot dimensions in the horizontal and vertical directions and a value of $\eta=1$ corresponds to the original, uncompressed snapshot.
In the present study, the snapshot dimensions for each flow example are set as $(m,n)=(500,300)$ for cylinder wake, $(128,128)$ for two-dimensional decaying turbulence, and $(256, 256)$ for turbulent channel flow, respectively.

\begin{figure}
    \centering
    \includegraphics[width=1\textwidth]{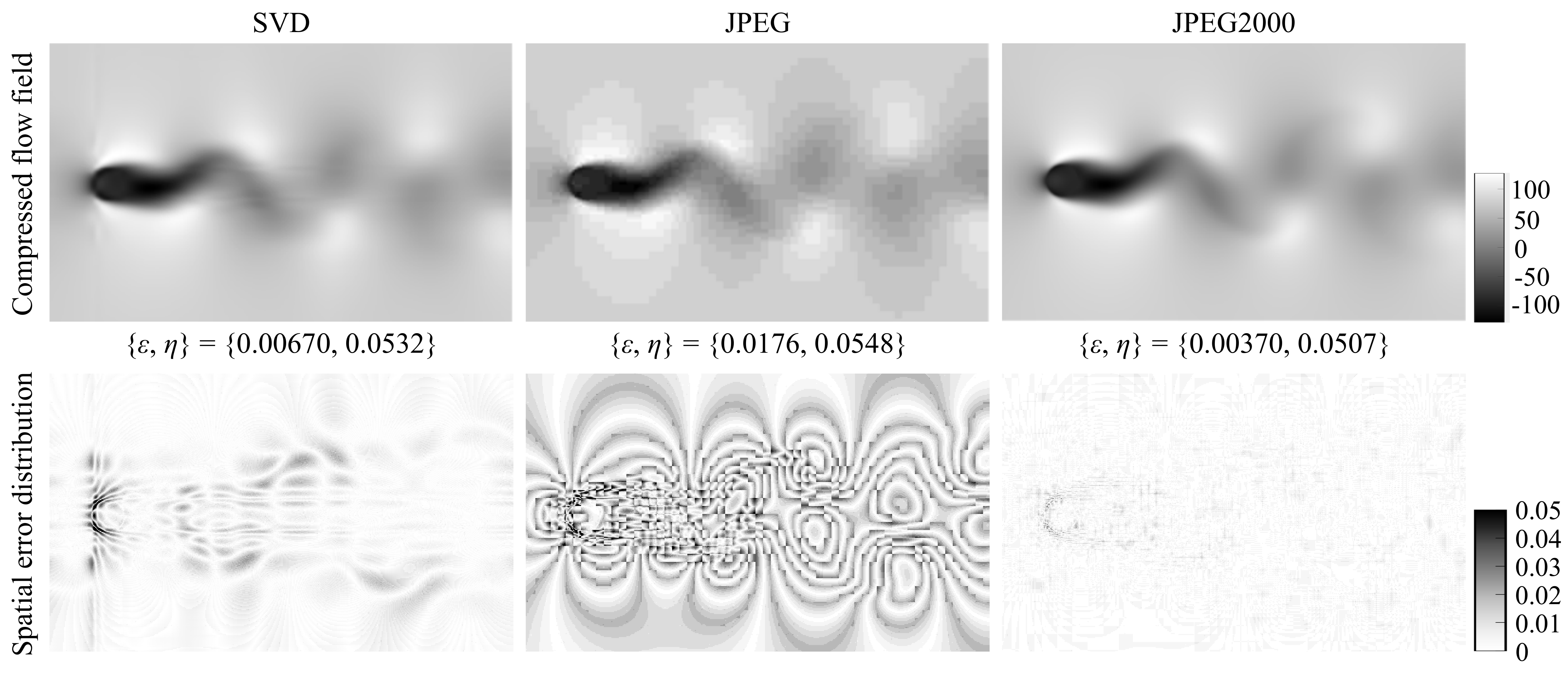}
    \caption{Comparison of image compression techniques for cylinder wake at $Re_D=100$.
    A streamwise velocity field $u$ is considered.
    The $L_2$ error norm of the reconstruction $\varepsilon$ and the compression ratio $\eta$ are shown underneath each flow field contour.
    Spatial absolute error distribution for each compression technique is also presented.
  }
    \label{fig:fowfieldComparison_a}
\end{figure}

Let us compare the image compression techniques with the cylinder wake example.
As for the data attribute, we use a streamwise velocity $u$.
The compressed wake fields and the spatial absolute error distributions with $\eta \approx 0.05$ are presented in figure~\ref{fig:fowfieldComparison_a}.
The $L_2$ error norm of reconstruction $\varepsilon = ||f_{\rm Ref}-f_{\rm Comp}||_2/||f_{\rm Ref}||_2$, where $f_{\rm Ref}$ and $f_{\rm Comp}$ are respectively the reference and compressed flow fields, is also shown underneath the decoded fields.
The SVD produces negligible error for the entire flow field, although slight discontinuities are observed in the wake region.
By comparison, JPEG compression introduces some compression artifacts including discontinuities of grayscale contours and granulated vortical structures.
This is due to fixed-size areas upon which the DCT is performed and elementary anti-blocking features. 
This indicates that compression based on the fixed block of $8\times 8$ pixels is not appropriate for flow fields that include fine-scale spatial variations.
In contrast, the compressed flow field with the JP2 algorithm retains wake features even while achieving significant data compression, with the $L_2$ error of 0.00370.
These results support the effectiveness of the adaptive block size of DWT in JP2 compression for bluff body wake data sets.

\begin{figure}
    \centering
    \includegraphics[width=1\textwidth]{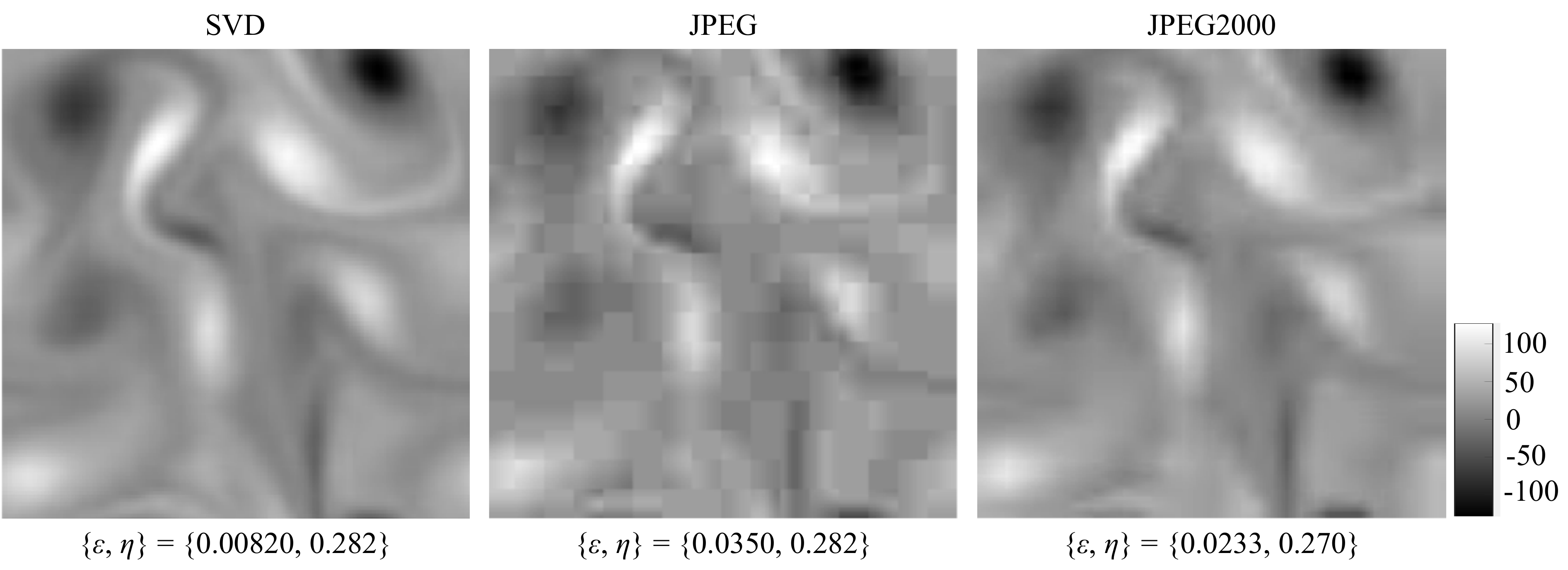}
    \caption{Comparison of image compression techniques for two-dimensional decaying turbulence.
    A vorticity field $\omega$ is considered.
    The $L_2$ error norm of the reconstruction $\varepsilon$ and the compression ratio $\eta$ are shown underneath each flow field contour.
  }
    \label{fig:fowfieldComparison_b}
\end{figure}

Next, we apply the image compression techniques to two-dimensional decaying homogeneous isotropic turbulence, as shown in figure~\ref{fig:fowfieldComparison_b}.
We use a vorticity field $\omega$ as a quantity of interest and compare the compression results with $\eta \approx 0.280$.
Similar to the cylinder example, the SVD can provide a smooth field and small error for the entire flow field.
Although SVD can achieve a reasonable compression for the laminar cylinder wake and two-dimensional turbulence that are mainly composed of large vortical structures, we discuss later how the presence of fine-scale turbulent structures alters the compression performance.
For this two-dimensional turbulence, the effect of $8\times 8$ pixel blocks can be clearly observed in JPEG compression.
Such pixelized artifacts on the flow field can be mitigated by using the JP2 compression technique, analogous to the observation with the cylinder example.

\begin{figure}
    \centering
    \includegraphics[width=1\textwidth]{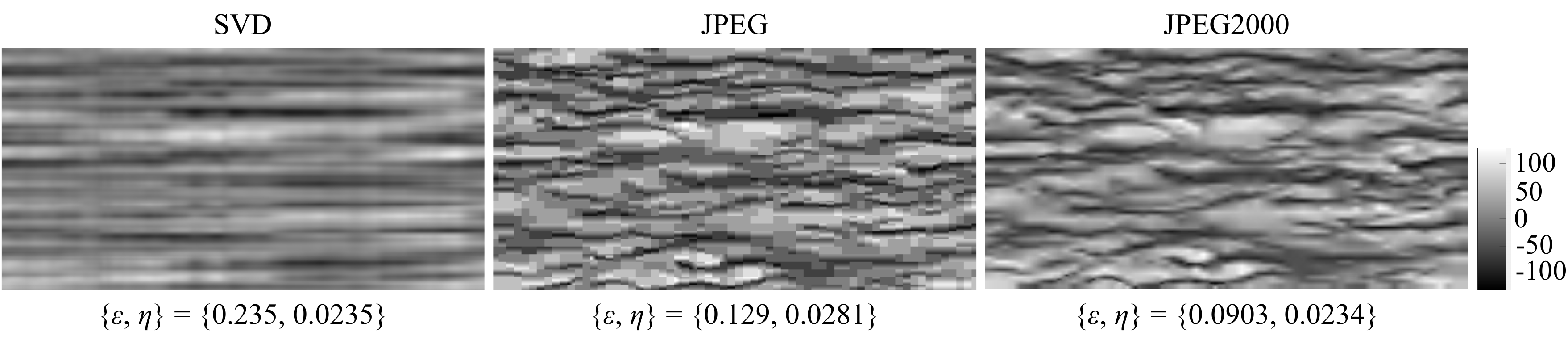}
    \caption{Comparison of image compression techniques for turbulent channel flow at $Re_\tau=180$.
    A streamwise velocity field $u$ is considered.
    The $L_2$ error norm of the reconstruction $\varepsilon$ and the compression ratio $\eta$ are shown underneath each flow field contour.
  }
    \label{fig:fowfieldComparison_c}
\end{figure}

The limitation of the SVD and the efficacy of the DWT-based process in the JP2 algorithm are further emphasized in the example of more complex turbulence.
Here, the compression techniques are applied to a streamwise velocity $u$ of the three-dimensional channel flow.
The compression results with $\eta \approx 0.025$ are compared in figure~\ref{fig:fowfieldComparison_c}.
As shown, the SVD-based compression cannot retain the important features of the streaks.
Compared to SVD, JPEG provides a better reconstruction although it also introduces discontinuities that obscure small spatial length scales in the flow field.
Surprisingly, JP2 produces non-negligible artifacts and maintains an $L_2$ error norm less than half that of SVD. 
The channel flow field at this low $\eta$ remains nearly indistinguishable from the uncompressed flow field, also preserving the streak spacing of the reference DNS field~\cite{KMM1987,smith1983characteristics}.
These observations suggest the effectiveness of the JP2 algorithm for image compression of complex fluid flow data.

\begin{figure}
    \centering
    \includegraphics[width=0.9\textwidth]{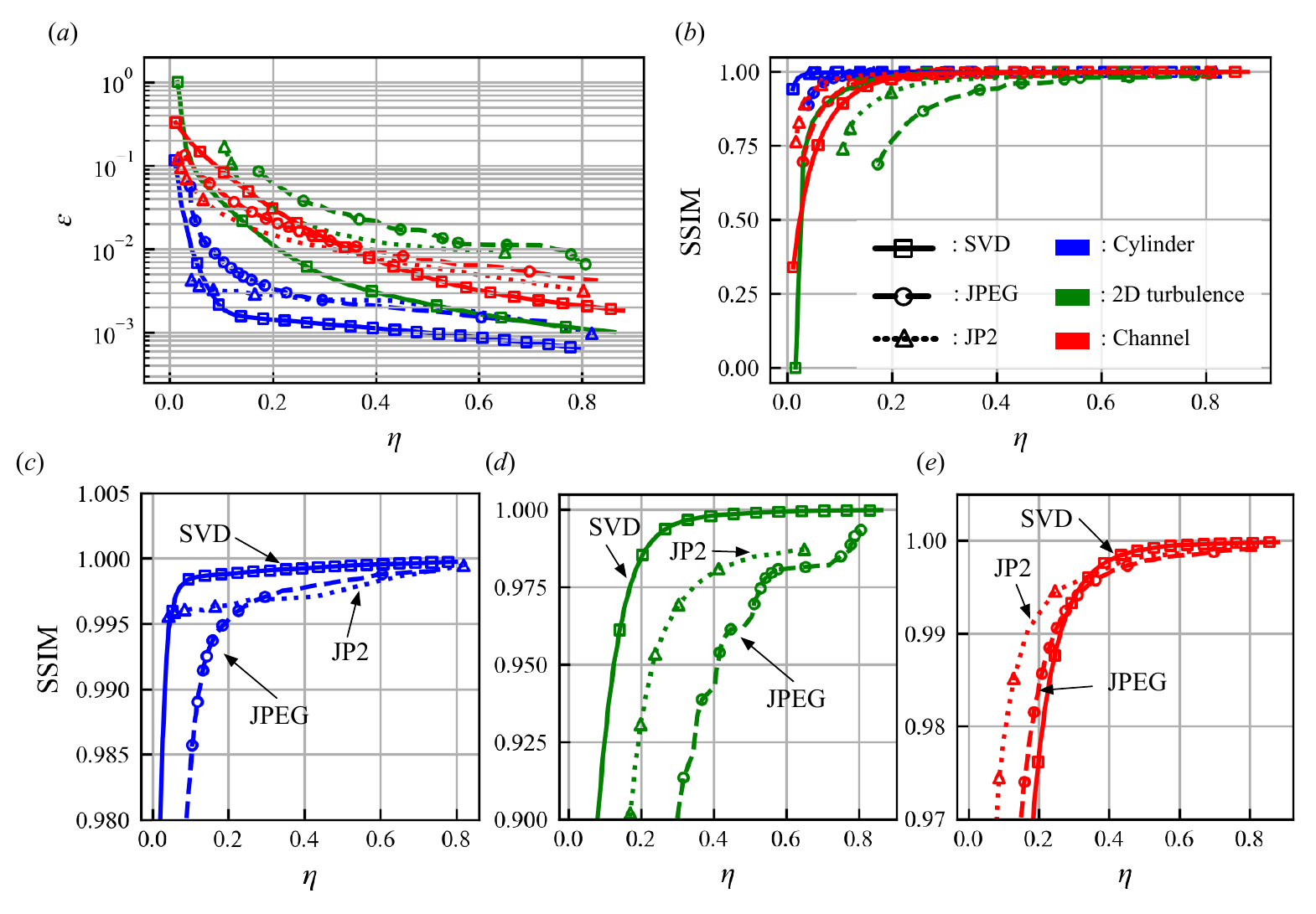}
  \caption{
    Relationship between $(a)$ the $L_2$ error norm $\varepsilon$, $(b)$ SSIM, and image compression ratio $\eta$.
    Zoom-in view of $\eta$-SSIM curve for $(c)$ cylinder wake, $(d)$ two-dimensional turbulence, and $(e)$ turbulent channel flow.
    }
\label{fig:SVDCompression}
\end{figure}

Building on these assessments, the $L_2$ error between compressed and uncompressed flow fields is evaluated across compression ratios, as shown in figure~\ref{fig:SVDCompression}$(a)$.
The error is averaged over all temporal snapshots of each flow example. 
In general, all compression algorithms produce an asymptotically decaying $L_2$ error.
JPEG introduces appreciable error at low $\eta$, in the same order as SVD compression.
It is worth pointing out that JP2 performs especially well at low $\eta$ while SVD compression produces the lowest $L_2$ error for high $\eta$ for all flow fields.

As an additional metric for quantifying the error introduced by each compression method, the localized structural similarity index (SSIM)~\cite{wang2004image} is computed between compressed and uncompressed flow fields.
SSIM can capture spatial correlation around pixels and is less sensitive against a pixel-wise error caused by translation and rotational difference compared to the $L_2$ error.
Hence, SSIM is suited for the image and video-based compression analysis. 
The SSIM $\chi$ is defined as
\begin{align}
    \chi = l(i_x,i_y)c(i_x,i_y)s(i_x,i_y)
\end{align}
where
\begin{equation}
    l(i_x,i_y) = \frac{2\mu_x \mu_y + C_1}{\mu_x^2 + \mu_y^2 + C_1},
    \quad
    c(i_x,i_y) = \frac{2\sigma_x \sigma_y + C_2}{\sigma_x^2 + \sigma_y^2 + C_2},
    \quad
    s(i_x,i_y) = \frac{\sigma_{xy} + C_3}{\sigma_x \sigma_y + C_3}
\end{equation}
with $\mu_x$ and $\sigma_x$ defined as the mean and standard deviation of $i_x$ respectively, $\sigma_{xy}$ being the covariance of $i_x$ and $i_y$, and $c_1$, $c_2$, and $c_3$ being constants to stabilize division. 
We set $\{C_1, C_2, C_3\} = \{0.16, 1.44, 0.72\}$ following Wang et al.~\cite{wang2004image}.
The resultant value lies between 0, representing no similarity, and 1, representing an identical image.
The relationship between the image compression ratio and the $L_2$ error is depicted in figure~\ref{fig:SVDCompression}$(b)$
Generally, JP2 and SVD produce a negligible decrease in the SSIM at low compression ratios and asymptotically approach an SSIM value of 1 at higher compression ratios. 
The SSIM value of the cylinder flow field with JPEG compression applied decays by approximately 10\%, as a result of significant discontinuities produced by JPEG.

We also present the zoom-in view of the relationship between SSIM and the compression ratio $\eta$ for each flow, in figures~\ref{fig:SVDCompression}$(c)-(e)$.
Similar to the observation in the $\eta-\varepsilon$ curves in figure~\ref{fig:SVDCompression}$(a)$, SVD and JP2 provide high SSIM scores compared to JPEG.
Especially at excessive compression (low $\eta$) of three-dimensional turbulent channel flow, JP2 can provide better reconstructions than the other two cases.
{Although scalar metrics such as the $L_2$ error $\varepsilon$ and SSIM are useful, we note that monitoring not only scalar values but also decoded flow fields with statistics is important in assessing how vortical structures can be retained through data compression because the influence of local structures are averaged.}

\begin{figure}
  \centerline{
  \includegraphics[width=\textwidth]{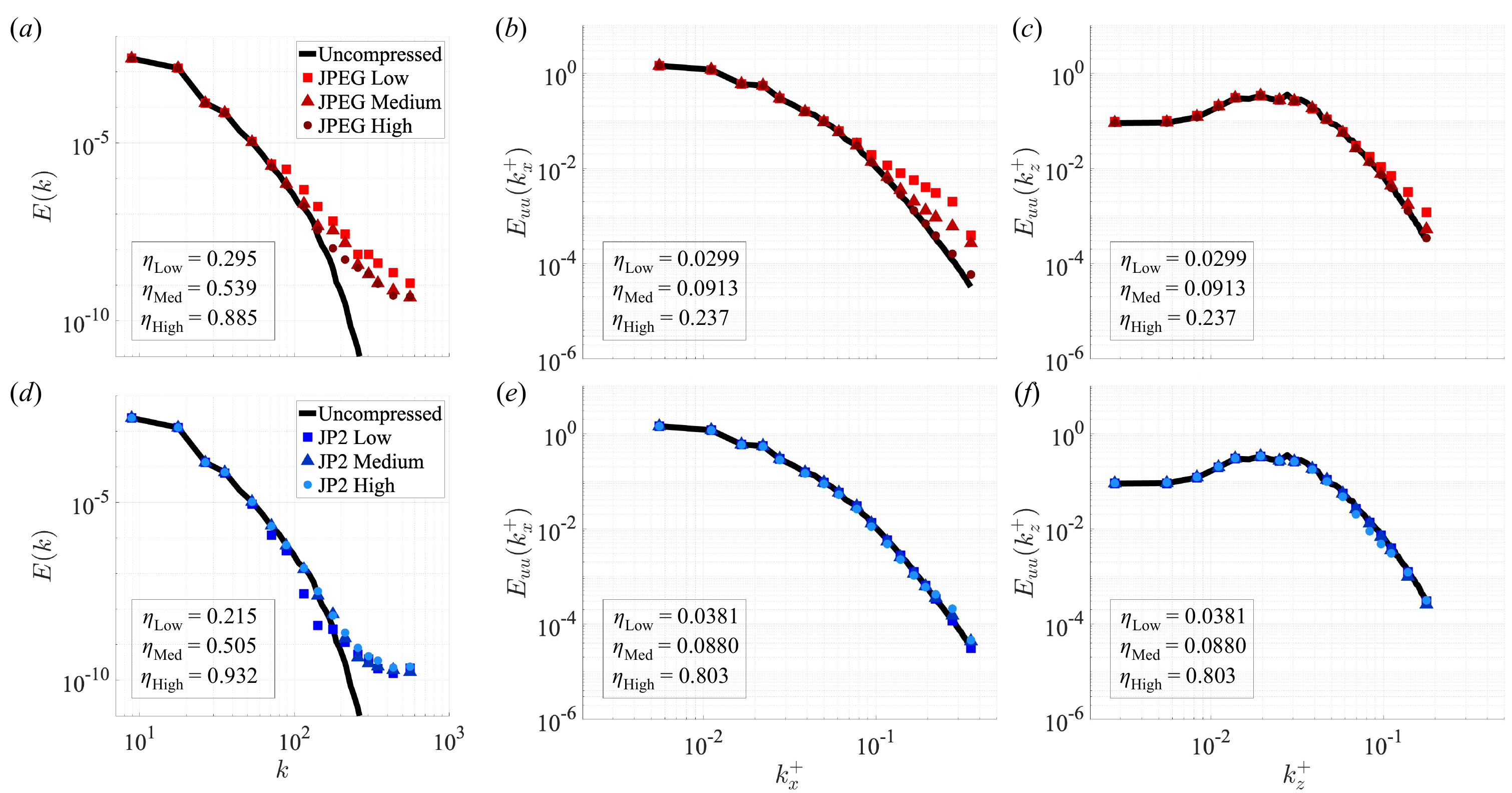}
  }
  \caption{
    Kinetic energy spectra for two-dimensional decaying homogeneous isotropic turbulence using $(a)$ JPEG and $(d)$ JP2.
    $(b,e)$ Streamwise and $(c,f)$ spanwise kinetic energy spectrum of three-dimensional turbulent channel flow compressed with $(b,c)$ JPEG and $(e,f)$ JP2.
}
\label{fig:kineticEnergy}
\end{figure}

We are additionally interested in whether finer structures in flow images can still be retained through the present compression process.
To examine this aspect, we consider the kinetic energy spectrum of both two- and three-dimensional turbulence examples, as summarized in figure~\ref{fig:kineticEnergy}.
The kinetic energy spectrum $E(k)$ for two-dimensional decaying turbulence is
\begin{align}
    E(k) = \dfrac{1}{2}(\overline{u_iu_i}),
\end{align}
where $u_i$ are the components of the fluctuating velocity and the overbar denotes an averaging operation in space and time.
For three-dimensional turbulent channel flow, the one-dimensional streamwise and spanwise spectra is evaluated
\begin{align}
    E_{uu}(k_x^+; y^+) = \overline{{\hat{u}}^\ast\hat{u}}^{z,t},~~E_{uu}(k_z^+; y^+) = \overline{{\hat{u}}^\ast\hat{u}}^{x,t},
\end{align}
where $(\cdot)^\ast$ represents the complex conjugate and $\hat{(\cdot)}$ denotes the one-dimensional Fourier transformed variable.
Here, we compare three compression ratios, denoted as low, medium, and high, for each turbulent flow.

JP2 demonstrates a strong adherence to the kinetic energy spectrum of the uncompressed flow field in both the $x$ and $z$ directions while JPEG compression at low $\eta$ introduces non-negligible errors at higher wave numbers. 
This is a consequence of the quantization step of JPEG compression that removes high wavelength scales from the image. Considering the overestimation of $E_{uu}(k^+_x)$ as seen in figure~\ref{fig:kineticEnergy} when using JPEG, this is likely caused by the absence of a deblocking filter, producing more high wavelength artifacts in the image than what exists in the uncompressed data. 
Similarly, the underestimation of $E(k)$ by JP2 can be attributed to adaptive block sizes that produce a lower peak signal-to-noise ratio, indicative of lower quality.
In general, JP2 is more adept at preserving high-wavenumber structures.

\subsection {Video compression}

\begin{figure}[t]
  \centering
  \includegraphics[width=0.8\textwidth]{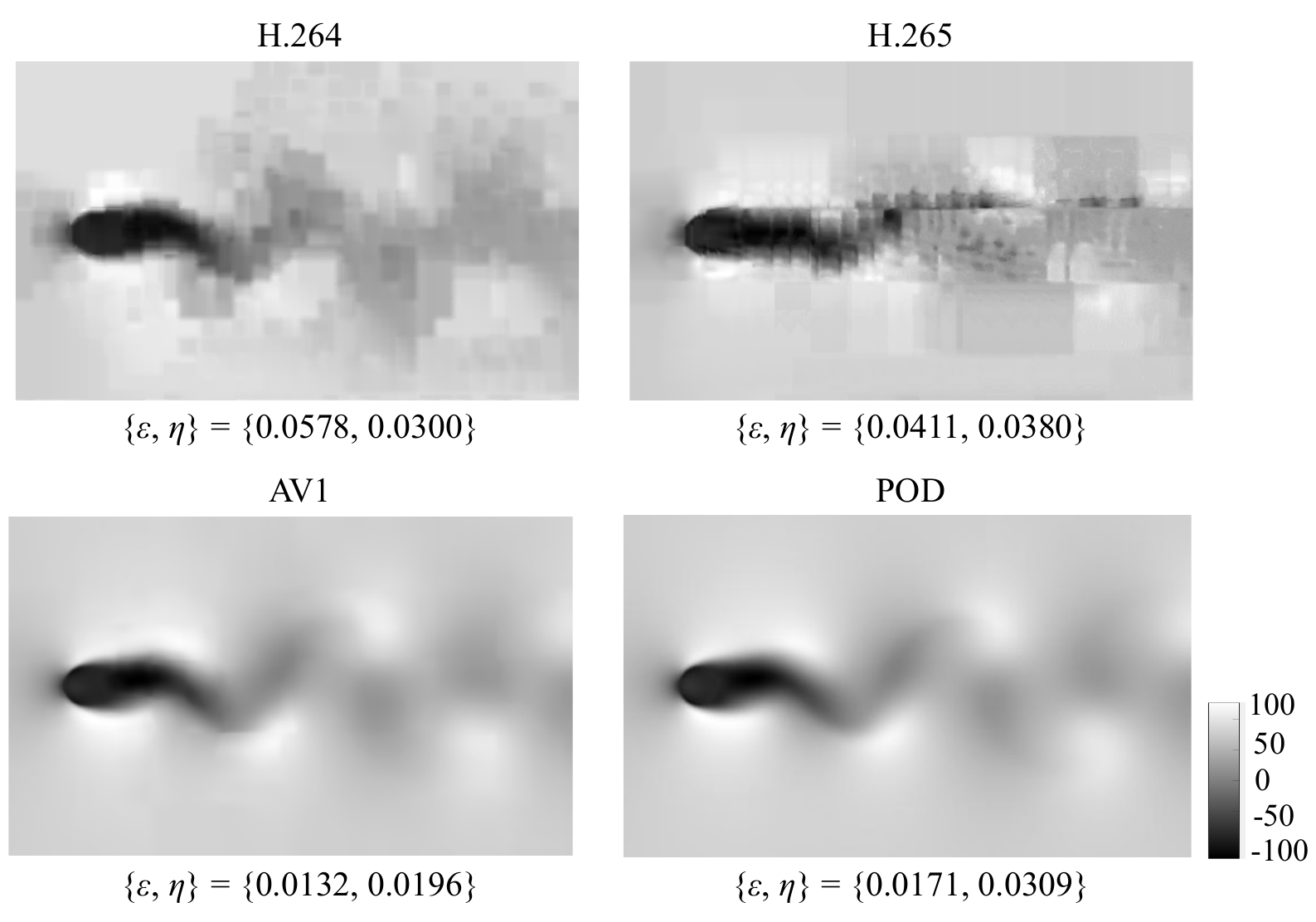}
  \caption{
    Comparison of video compression techniques applied on a streamwise velocity field $u$ of cylinder wake at $Re_D=100$, compressed using H.264, H.265, AV1, and POD compression algorithms.
    The $L_2$ error norm of the reconstruction $\varepsilon$ and the compression ratio $\eta$ are shown underneath each flow field contour.
  }
\label{fig:videoCompressionOverview1}
\end{figure}

\begin{figure}[h]
  \centering
  \includegraphics[width=1\textwidth]{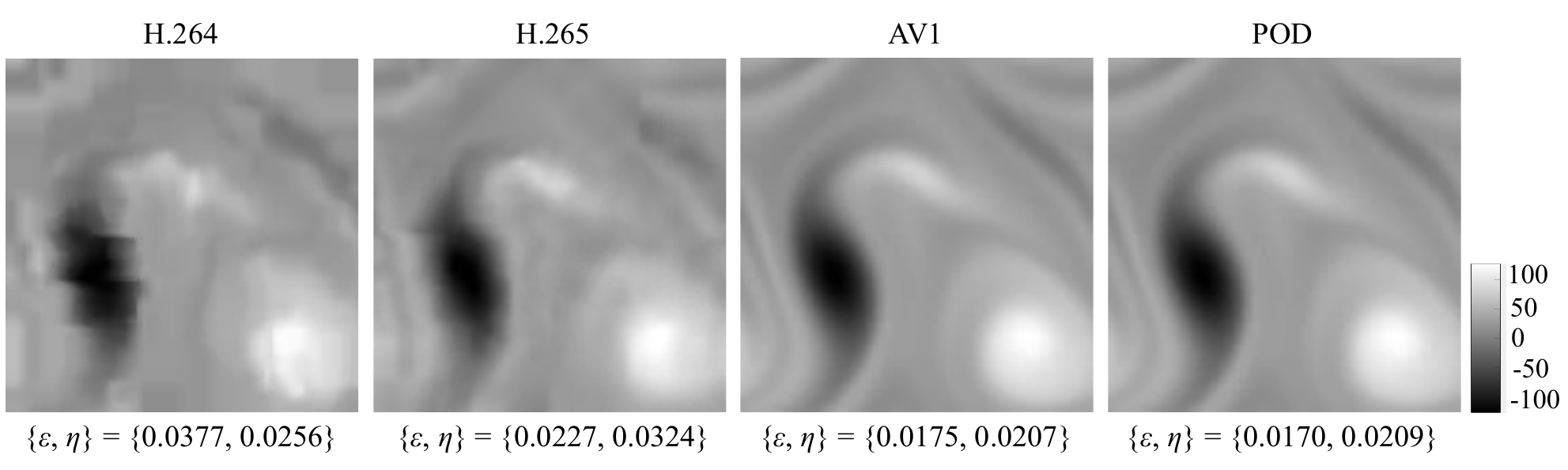}
  \caption{
    Comparison of video compression techniques applied on two-dimensional isotropic turbulent vorticity field, compressed using H.264, H.265, AV1, and POD compression algorithms.
    The $L_2$ error norm of the reconstruction $\varepsilon$ and the compression ratio $\eta$ are shown underneath each flow field contour.
  }
\label{fig:videoCompressionOverview2}
\end{figure}

\begin{figure}
  \centering
  \includegraphics[width=0.85\textwidth]{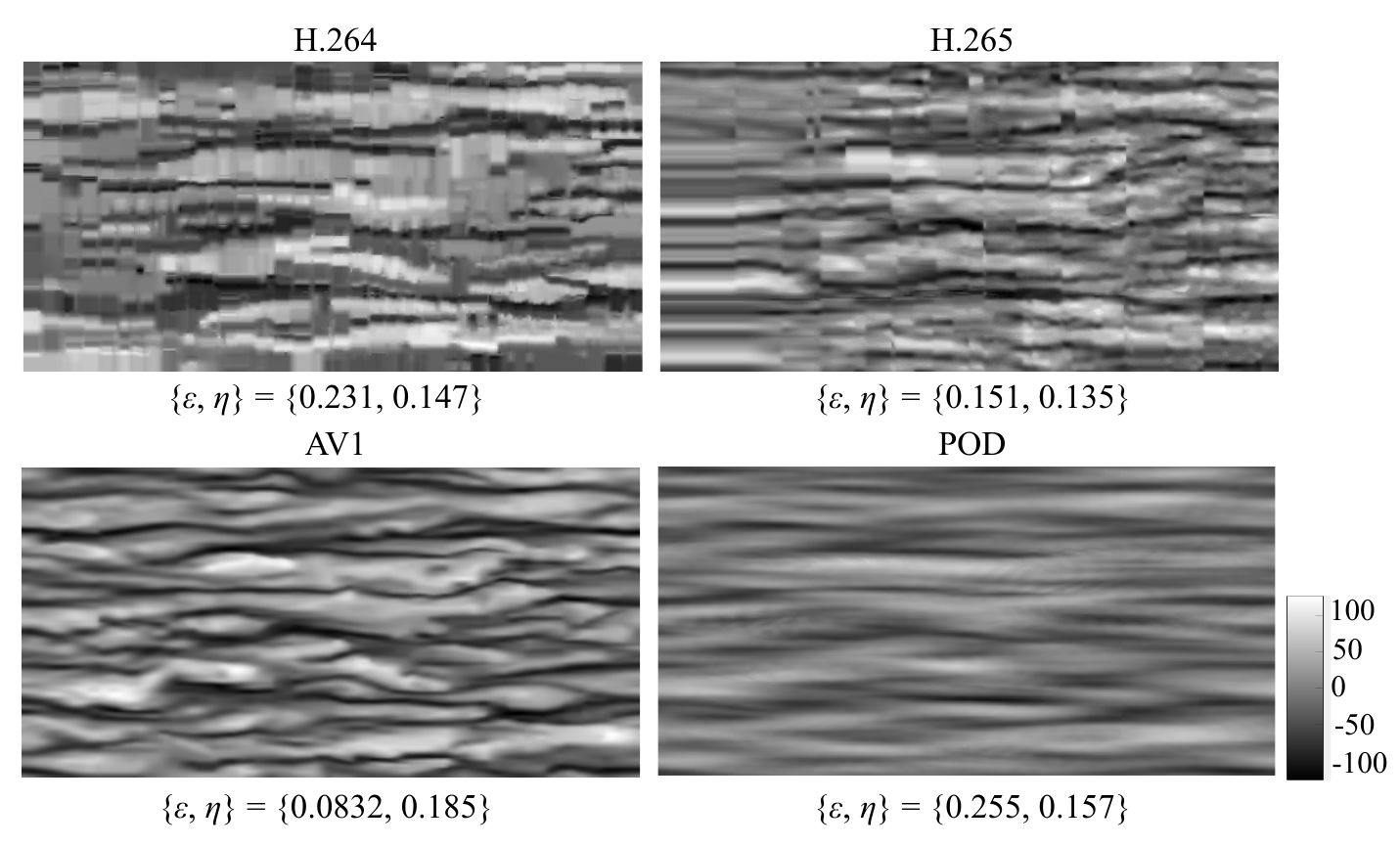}
  \caption{
    Comparison of video compression techniques applied on a streamwise velocity field $u$ of three-dimensional turbulent channel flow at $Re_\tau=180$, compressed using H.264, H.265, AV1, and POD compression algorithms.  
    The $L_2$ error norm of the reconstruction $\varepsilon$ and the compression ratio $\eta$ are shown underneath each flow field contour.
  }
\label{fig:videoCompressionOverview3}
\end{figure}

From the perspective of information, fluid flows are inherently temporally-redundant — as such, video compression algorithms that perform temporal compression are a powerful tool, achieving compression performance that outperforms the previously analyzed image-based techniques. 
This section assesses the capabilities of video compression techniques such as H.264, H.265, and AV1 compression algorithms for time-varying fluid flow data.
Additionally, proper orthogonal decomposition (POD) compression~\cite{Lumely1967} is considered to compare this familiar method of compression within the fluid dynamics community with those analyzed herein~\cite{taira17,THBSDBDY2020}.
POD is used to decompose a matrix of vectorized, temporally evolving flow field data into a set of basis modes and eigenvalues that contain coherent flow structures and can be used for flow field reconstruction. 
Formally, a flow field $\bm{q}(\bm{x},t) - \overline{\bm{q}(\bm{x})}$ can be represented as $\sum_{j=1}^n a_j\bm{\phi_j}$ where $a_j$ is the temporal coefficient for mode~$\bm{\phi_j}$. 
The value of $a_j$ is the inner product between the mode $\bm{\phi_j}$ and the mean-subtracted flow field, $\bm{q}(\bm{x},t) - \overline{\bm{q}(\bm{x})}$.
This modal representation can be truncated to $r$ modes, such that the flow field is approximated by $\sum_{j=1}^r a_j\bm{\phi_j}$.
This study uses the snapshot POD method~\cite{sirovich1987turbulence} for comparison to the other video compression techniques.
The compression ratio for a reconstructed flow field containing $r$ modes is evaluated as 
\begin{align}
    \eta = \dfrac{r(m+n)+m}{n(m+n)+m},
\end{align}
where $m$ is the total number of pixels in the flow field and $n$ is the total number of flow snapshots. 
{Note that POD is hereafter used for time-series of flow snapshots in comparing to the video compression techniques while we performed an instantaneous SVD for image compression, }

The results of video compression for laminar cylinder wake at $Re_D=100$ are shown in figure~\ref{fig:videoCompressionOverview1}.
Here, we compare the decoded streamwise velocity field $u$ with $\eta \approx 0.02-0.03$.
POD compression introduces negligible error, likely as a result of the temporally redundant nature of periodic wake and the larger coherent modal structures that POD is able to extract.
By comparison, H.264 compression produces significant artifacts at low $\eta$. 
H.264 struggles likely because inter-frame prediction candidates are chosen from a shallow time range. 
We also observe that H.265 fails to improve in terms of error level over H.264 for the cylinder wake. 
This is due to the employment of a similar inter-frame prediction and selection algorithm to that of H.264. 
Compared to these H.2XX series, the AV1 algorithm provides much better compression, achieving a lower $L_2$ error than that achieved by H.264 and H.265. 
This highlights the enhanced capability of AV1 to compress laminar and temporally redundant flow fields.

We next examine the video compression techniques for two-dimensional decaying homogeneous isotropic turbulence, as summarized in figure~\ref{fig:videoCompressionOverview2}.
The flow fields are compared for the compression ratios of $\eta \approx 0.02-0.03$.
Similar to the cylinder case, POD compression provides a reasonable reconstruction, likely because large-scale vortical structures are dominant at this particular time.
It is, however, easily anticipated that the error of this time-varying flow relies on the presence of a range of length scales, as the small length scales disappear with the progress of the decay over time~\cite{mcwilliams1984emergence,yeh2021network}.
The dependence of the compression performance over time for decaying flow will be examined later.
While H.2xx compression techniques provide a reasonable reconstruction, AV1 provides better compression without suffering from pixelized artifacts.
These results suggest the powerful capabilities of novel deblocking filters for fluid flow applications.

The video compression techniques are also applied to the $x-z$ sectional streamwise velocity field $u$ of three-dimensional turbulent channel flow at $Re_\tau = 180$, as depicted in figure~\ref{fig:videoCompressionOverview3}.
The compressed flow fields are compared for $\eta \approx 0.150$. 
In contrast to the other flow examples, POD compression produces significant visible artifacts and a high error value for turbulent channel flow because of a complex temporal evolution of the flow field.
POD requires a greater number of modes for adequate reconstruction~\cite{alfonsi2006,MPMF2019,fukami2020convolutional}.
Although H.265 improves over H.264 significantly for turbulent channel flow, this still produces few observable discontinuities.
This is likely caused by adaptive tiling in macroblocks for prediction procedures, allowing lower $\eta$ with similar flow field representation. 
AV1 exceeds the performance of H.264 and H.265 consistently and POD compression on turbulent channel flow.
Flow fields compressed using AV1 are indistinguishable from uncompressed flow fields at high $\eta$.

\begin{figure}
  \centerline{
  \includegraphics[width=0.85\textwidth]{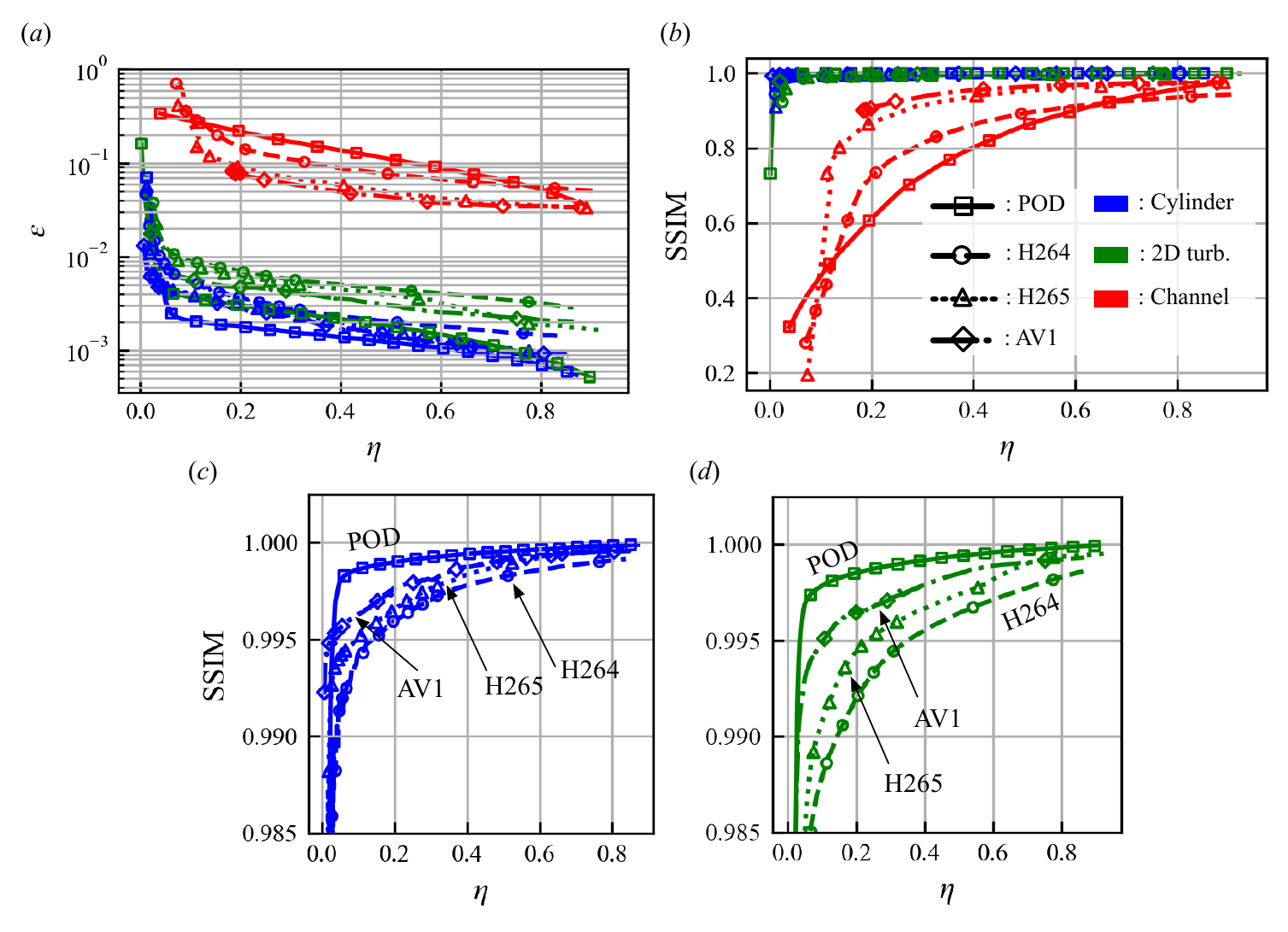}
  }
  \caption{
    Relationship between $(a)$ the $L_2$ error norm $\varepsilon$, $(b)$ SSIM, and video compression ratio $\eta$.
    Zoom-in view of $\eta$-SSIM curve for $(c)$ cylinder wake and $(d)$ two-dimensional isotropic turbulence.}
\label{fig:videoCompressionError}
\end{figure}

The $L_2$ error and SSIM are evaluated across a range of compression ratios for each type of flow field, as presented in figure~\ref{fig:videoCompressionError}. 
In general, all video compression algorithms produce asymptotically decaying $L_2$ error values with increasing $\eta$.
AV1 performs well at low $\eta$, especially for the cylinder wake. 
Additionally, all compression algorithms perform well for two-dimensional turbulence, likely as a result of the flow field snapshots holding slow changes from one frame to the next due to the decaying nature of the flow.
Moreover, as observed with samples at various compression ratios, the $L_2$ error for all algorithms plateau at non-negligible values for the cylinder wake flow field. 
SSIM values generally diverge from the asymptotic limit at low $\eta$.
The exceptional cases include cylinder wake and two-dimensional turbulence compressed using AV1, which introduces negligible error at low $\eta$. 
AV1 outperforms H.264 and H.265 on turbulent channel flow as well, due to the improved blocking techniques.

\begin{figure}
  \centering
  \includegraphics[width=\textwidth]{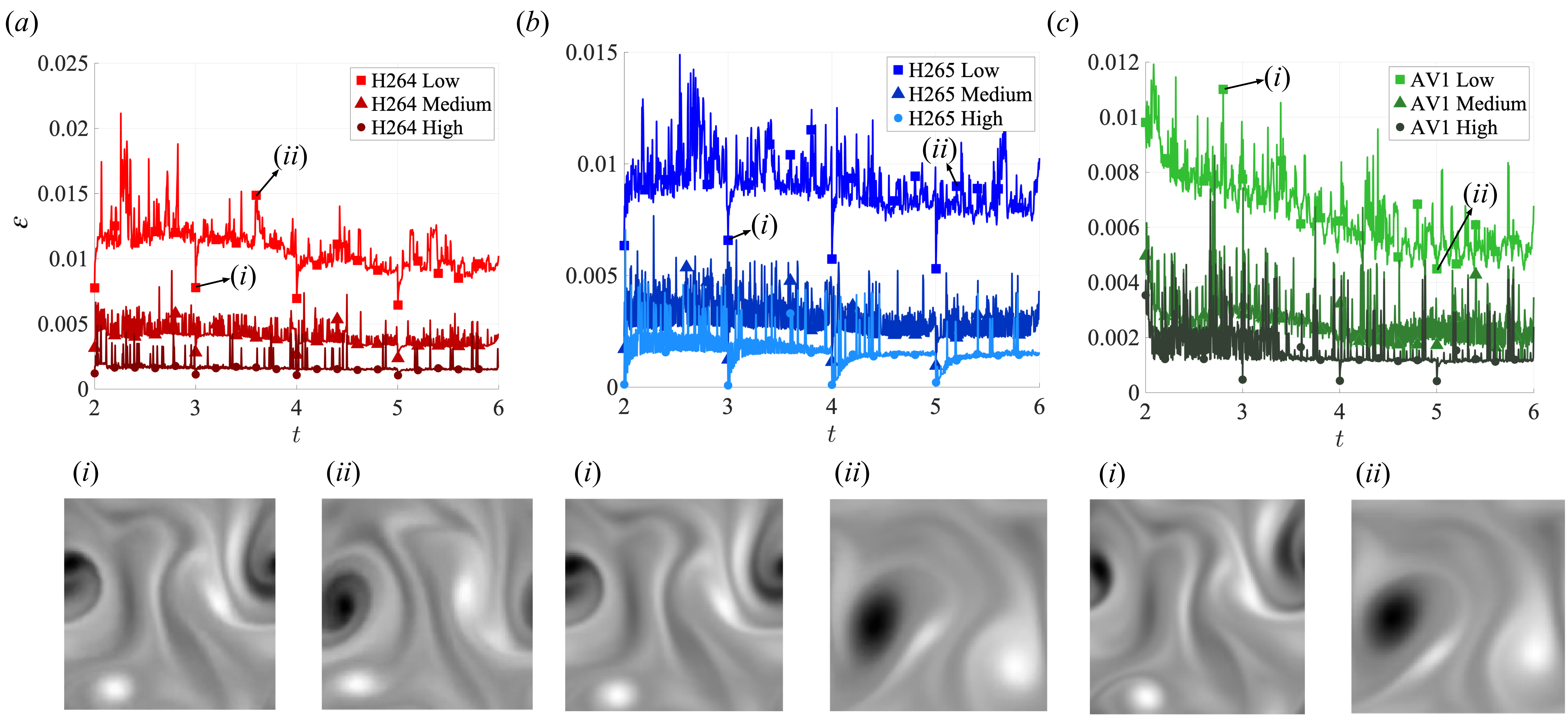}
    \caption{
    $L_2$ error norm $\varepsilon$ of vorticity field $\omega$ for two-dimensional decaying homogeneous isotropic turbulence over time. 
    $(a)$ H.264, $(b)$ H.265, and $(c)$ AV1.
    $(i)$ and $(ii)$ in each case are chosen due to their employment in inter-frame prediction in each algorithm.
  }
\label{fig:videoCompressionL2OverTime_a}
\end{figure}

In addition, the time evolution of the $L_2$ error is examined to gain insight into the performance of video compression algorithms for individual snapshots. 
The temporal evolution of the $L_2$ error norm $\varepsilon$ for two-dimensional decaying turbulence is shown in figure~\ref{fig:videoCompressionL2OverTime_a}.
H.2xx compression techniques exhibit repeated temporal structures in its $L_2$ error evolution, likely as a consequence of inter-frame prediction selecting frames to make predictions from at relatively similar intervals. 
AV1 compression provides a distinctive reduction in the $L_2$ error over time for medium and low $\eta$, indicating improved accuracy as snapshots begin to show redundancies due to the vortex field decaying and exhibiting similar large-scale coherent structures from one snapshot to the next.
We also observe that H.264 produces a high error at low $\eta$ for early flow field snapshots. 
This relates to the time-varying flow nature of the present decaying turbulence, as mentioned above.
The presence of finer structures at the high Taylor Reynolds number $Re_\lambda(t)$ portion of the flow likely causes the difficulty in compressing vortical flow data.

\begin{figure}
  \centering
  \includegraphics[width=\textwidth]{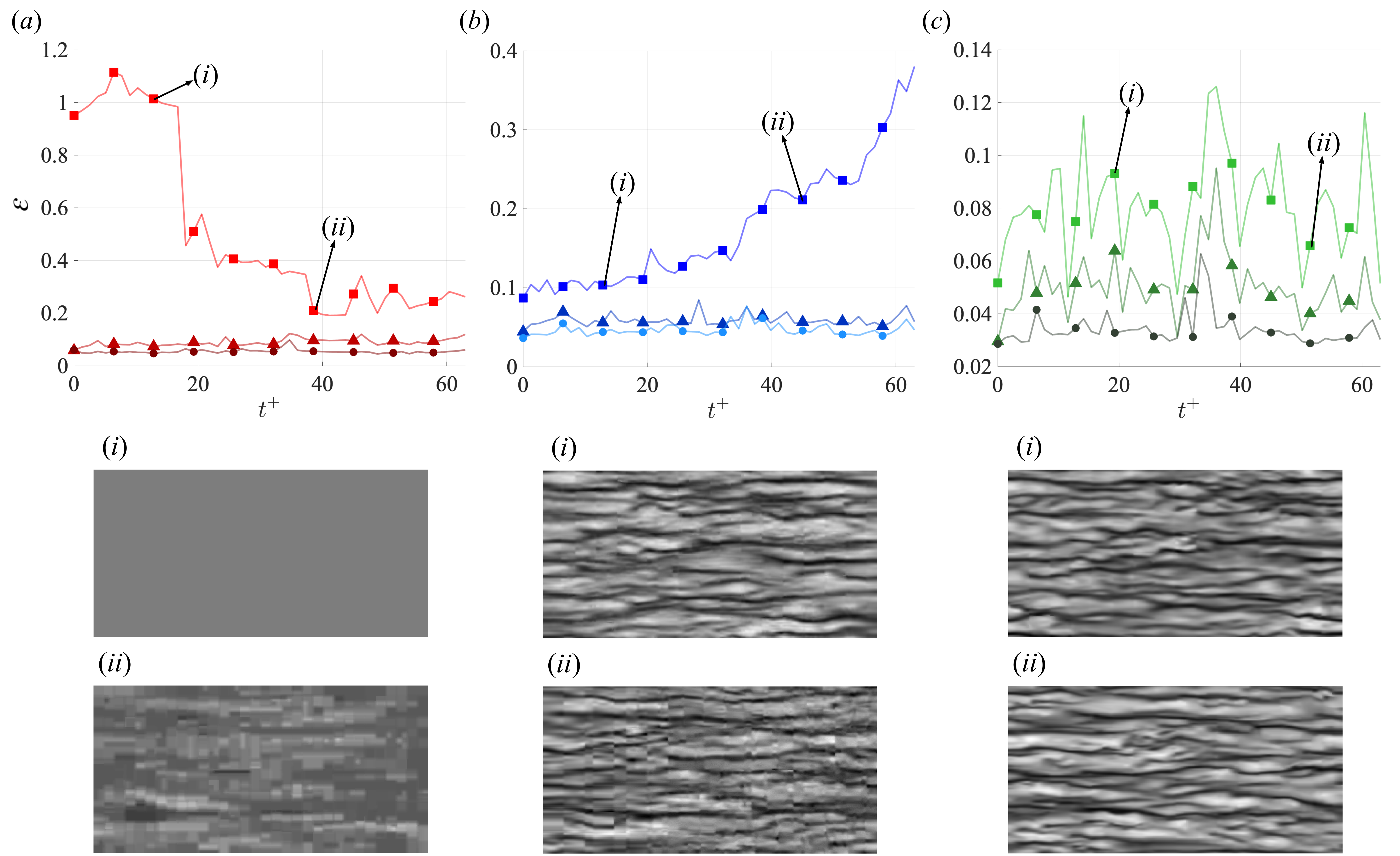}
  \caption{
    $L_2$ error norm $\varepsilon$ of streamwise velocity field $u$ for turbulent channel flow over time. 
    $(a)$ H.264, $(b)$ H.265, and $(c)$ AV1.
    $(i)$ and $(ii)$ in each case are chosen due to their employment in inter-frame prediction in each algorithm.  }
\label{fig:videoCompressionL2OverTime_b}
\end{figure}

We also examine the $L_2$ error norm $\varepsilon$ and the flow fields over time for turbulent channel flow, as depicted in figure~\ref{fig:videoCompressionL2OverTime_b}.
H.264 generally produces a larger $L_2$ error compared to the other techniques, as we also observed with the visual assessments in figure~\ref{fig:videoCompressionOverview3}.
With low $\eta$ of H.264 compression, the error decreases over time, likely as a result of a later snapshot being selected for inter-frame prediction. 
Compared to H.264, H.265 provides better compression over time.
Similar to the observation with H.264, the $L_2$ error significantly varies over time at a low $\eta$.
This is likely due to the inter-frame selection of an early frame from which further predictions were made. 
AV1 produces a negligible error at a high $\eta$ while the errors increase as $\eta$ decreases.

\begin{figure}[t]
  \centerline{
  \includegraphics[width=\textwidth]{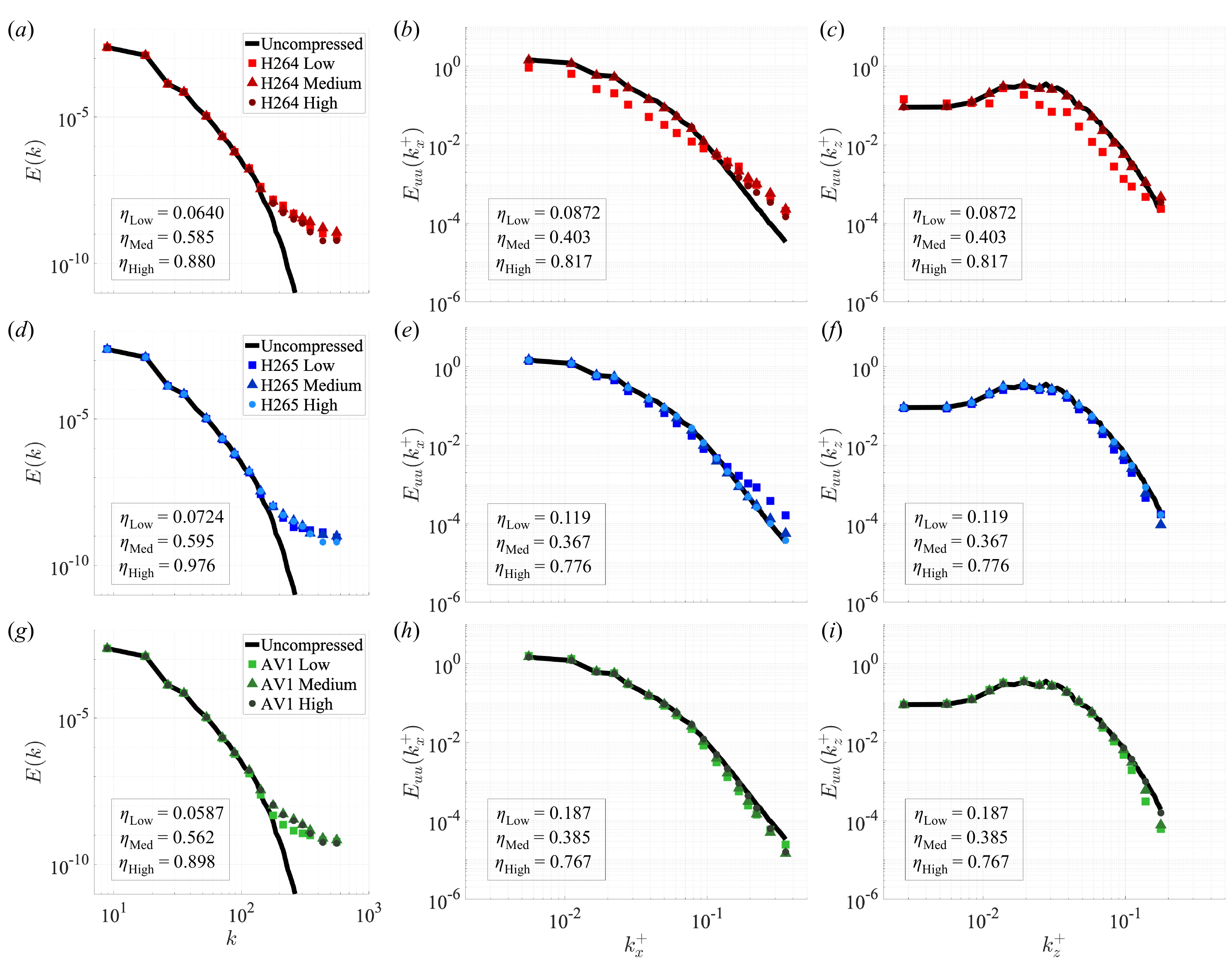}
  }
  \caption{
    Kinetic energy spectra $E(k)$ for $(a,d,g)$ two-dimensional decaying homogeneous isotropic turbulence using H.264, H.265, and AV1.
    $(b,e,h)$ Streamwise $E_{uu}(k_x^+)$ and $(c,f,i)$ spanwise kinetic energy spectra $E_{uu}(k_z^+)$ of three-dimensional turbulent channel flow.
  }
\label{fig:videoCompressionKE}
\end{figure}

We are also interested in the performance of video compression algorithms in preserving high wavenumber structures in the compressed state. 
The general performance of each compression algorithm with regard to kinetic energy spectra of each flow field is investigated, as shown in figure~\ref{fig:videoCompressionKE}.
H.264 performs well for two-dimensional turbulence, but produces a noticeable error at all $\eta$ in both the stream- and spanwise directions of the kinetic energy spectrum of turbulent channel flow.
A similar divergence from the expected data can be observed at low $\eta$ in the spanwise direction as well. 
H.265 performs comparatively well for two-dimensional decaying isotropic turbulence, and for turbulent channel flow in the spanwise direction.
However, it produces a non-negligible error at high wavenumbers when compressed at low $\eta$ in the spanwise direction.
This indicates an over-representation of high-wavenumber components due to blocking as a result of the adaptive subblock sizes of H.265.
Generally, AV1 is the best-suited algorithm for preserving spatial frequency information, particularly at high wavenumbers, for both two and three-dimensional turbulent flow fields. 
At higher $\eta$, the energy contents at each wavenumber are almost indistinguishable.

\begin{figure}
  \centerline{
  \includegraphics[width=\textwidth]{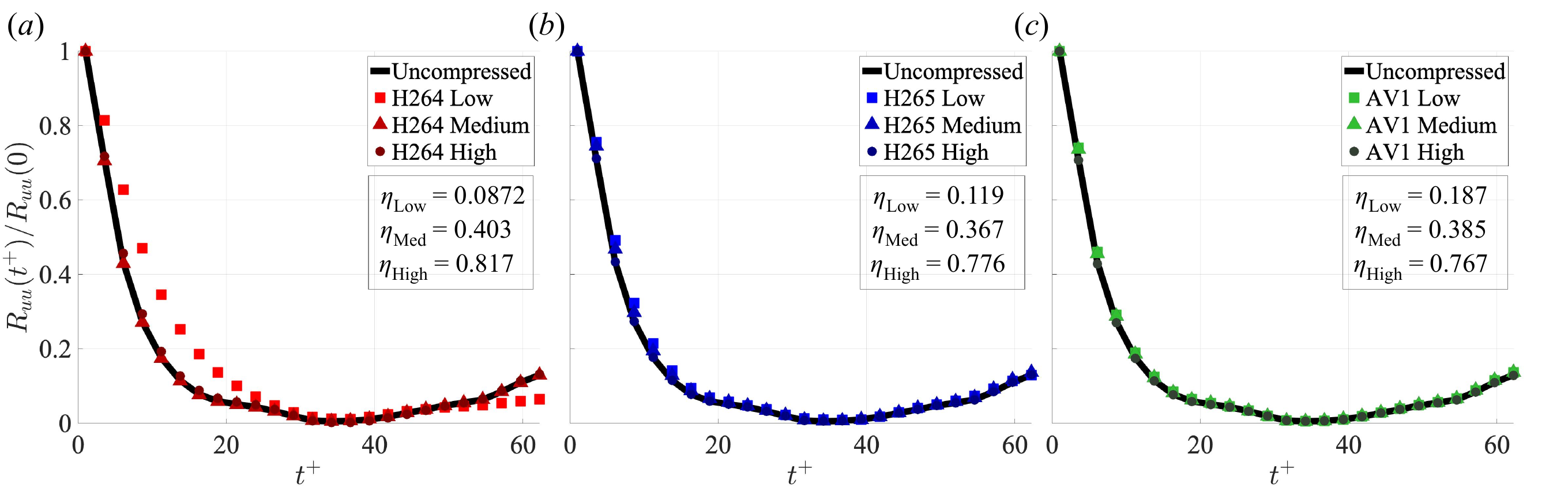}
  }
  \caption{
    Normalized temporal two-point correlation coefficients $R_{uu}(t^+)/R_{uu}(0)$ for three-dimensional turbulent channel flow using H.264, H.265, and AV1.
  }
\label{fig:videoCompressionTemp2Point}
\end{figure}

At last, we investigate whether the video compression techniques can preserve the temporal evolution of complex turbulent flows.
Here, let us examine the temporal two-point correlation for three-dimensional channel flow compressed using all three video compression algorithms. 
The temporal two-point correlation coefficient at a given $t^+$ is defined as
$R_{uu}^+(t^+)/R_{uu}^+(0)$~\cite{PRF2019,quadrio2003integral} and is depicted in figure~\ref{fig:videoCompressionTemp2Point}.
The assessment of temporal two-point correlation provides insight into the relations of flow snapshots to preceding snapshots.

Consistent with the insights gained from the kinetic energy spectrum, H.264 compression at a $\eta$ exhibits disagreement with the reference curve at $t^+$ values between 5 and 30, and above 50.
This is indicative of a de-correlation of the velocity field and is likely a result of poor performance in capturing high-wavenumber information.
Except for this particular case, all compression algorithms generally perform well, with temporal two-point correlation coefficients closely following that of the uncompressed flow field. 
These results suggest that these novel video compression techniques capture the spatio-temporal redundancies well even for complex turbulent flows and also significantly reduce data size while preserving their physics.

\section{Conclusion}
\label{sec:types_paper}

We compressed flow field data from canonical flow examples using a number of widely-available multimedia compression techniques. The performance of the JPEG and JP2 spatial image compression techniques and the H.264, H.265, and AV1 spatio-temporal video compression techniques were considered for simulated laminar cylinder flow, decaying isotropic turbulence, and turbulent channel flow.
Streamwise velocity and vorticity field data were represented as grayscale images and videos, and were compressed using the aforementioned techniques.

All techniques, with the exception of JPEG, were shown to compress flow data below 10\% of the original file size while introducing negligible error and preserving underlying flow physics. 
AV1 and H.265 compression were shown to have the best performance across a variety of flow regimes. 
The spatial error distributions were concentrated on the cylinder surface and directly behind the cylinder for the streamwise velocity data compression and in the vortex shedding wake for the vorticity data. 
Turbulence statistics in the form of kinetic energy spectra were preserved under compression for all methods except JPEG.

For single snapshots of data represented as an image, JP2 compression was shown to far outperform JPEG compression, with a tolerable increase in computational complexity. 
For multiple temporal snapshots of data represented as a video, the choice of compression method becomes more nuanced.
JP2 compression was shown to achieve the lowest compression error as temporal compression adds slight error to the data. 
The AV1 algorithm maximizes $\eta$ at the expense of computational complexity and non-negligible encoding time. 
This algorithm is new and emerging from the research environment, so future optimizations could bring this encoding time to a manageable level. 
The H.265 algorithm provided excellent compression performance at a fast encoding time, and appears as a promising algorithm for current fluid dynamics applications. 
H.264 provided acceptable compression performance, but was largely triumphed by the AV1 and H.265 algorithms.
{AV1 exhibited a significant difference of compression performance compared to conventional techniques for complex turbulence data.
Hence, it can be argued that AV1 can be especially recommended in compressing complex turbulent flows.
}

We have shown that modern multimedia compression algorithms provide robust performance in a variety of fluid flow applications. 
The implementation of these techniques becomes especially pertinent as simulations within computational fluid dynamics become exceedingly data-intensive, a trend that decreases the accessibility to high-fidelity models. 
These methods are free, easily accessible, regularly updated and supported, and provide flexible and scalable compression performance. 
{Multimedia compression can also support fast data transfer from open fluid-flow databases~\cite{li2008public,wu2008direct,towne2022database}, promoting data-based analysis in fluid mechanics.}
As such, the implementation of these compression techniques has exciting potential across the fluid dynamics community for data storage and transfer with minimal loss.

\section*{Acknowledgements}

KT acknowledges the support from the US Army Research Office (W911NF-21-1-0060), the US Air Force Office of Scientific Research (FA9550-21-1-0178), and the US Department of Defense Vannevar Bush Faculty Fellowship (N00014-22-1-2798). 
We also thank Professor Koji Fukagata (Keio University) for sharing his DNS code.

\section*{Appendix: Encoding time}

The increased performance of new compression algorithms comes at a cost; non-negligible increases in computational complexity should be considered when implementing these algorithms. 
In fact, in a paper from 2000 on compressing three-dimensional flows with the JPEG and JP2 algorithms~\cite{schmalzl03}, the added complexity of the JP2 algorithm caused JPEG to be recommended over JP2, despite losing clear performance benefits. 
The recommendation of the present study reverses that statement.
As such, it is important to quantify the encoding time of these algorithms at the time of writing this study.

The decoding time is observed to be negligibly small for all compression codecs; thus, this appendix focuses on encoding. 
The streamwise velocity and vorticity data are encoded for both the laminar cylinder flow and turbulent channel flow cases at the same bitrate (100 KB/s) for all compression algorithms and the encoding time is measured.
The encoding is performed with a 2.5GHz i7 Intel Core processor and 8 GB RAM.
The results are summarized in table \ref{tab:time}.
Encoding time per frame is observed to be larger for the turbulent channel flow than the laminar cylinder flow, indicating that the algorithms struggle to encode multiscale turbulent flow data. 
Across encoding algorithms, JPEG and H.264 compression are the fastest, a testament to the maturity and low complexity of these methods. 
JP2 and H.265 encoding are generally several times slower, but still relatively fast, justifying their added compression performance. 
AV1 is observed to be far slower in encoding than the other methods: over 100 times slower than JPEG and H.264, and over 25 times slower than JP2 and H.265. 
This severe encoding time increase limits the practicality of implementing this algorithm in large-scale applications, and perhaps justifies the use of H.265 over AV1.
As the algorithm was released only a few years prior to the writing of this paper, advances in computing and algorithm development could increase its practicality in the near future.

\begin{table}
\begin{center}
\def~{\hphantom{0}}
  \begin{tabular}{ c | c | c | c | c | c }
                               & JPEG  & JP2  & H.264 & H.265 & AV1   \\
     \hline
    $u$, Cylinder Flow         & 0.58  & 2.27 & 0.94  & 2.85  & 61.21 \\
    $\omega$, Cylinder Flow    & 0.59  & 1.53 & 0.92  & 2.39  & 43.83 \\
   $u$, Channel Flow           & 0.29  & 1.42 & 0.45  & 1.96  & 49.59 \\
  \end{tabular}
  \caption{Encoding time (s) for different compression algorithms and flow regimes, compressed at 100 KB/s bitrate.}
  \label{tab:time}
\end{center}
\end{table}

\section*{Declarations}

\subsection*{Conflict of interest}
The authors declare that they have no conflict of interest.
 
\subsection*{Authors' contributions}
KT designed research. 
VA, JF, and KF performed research and analyzed data. 
VA, JF, KF and KT wrote the paper. KT supervised.

\subsection*{Availability of data and materials }
The data that support the findings of this study are available from the corresponding author upon reasonable request.


\bibliographystyle{unsrt}  
\bibliography{arxiv}

\end{document}